\documentclass[english]{tlp}

\usepackage[latin1]{inputenc}
\usepackage[T1]{fontenc}
\usepackage{babel}
\usepackage{graphicx}
\usepackage{amsmath}
\usepackage{amsfonts}
\usepackage{mathtools}  % Mau: per il comando \begin{matrix*}
\usepackage{amssymb}    %
\usepackage{multirow}
\usepackage{multicol}
\usepackage{breakcites}
\usepackage{url}
\usepackage{stmaryrd}
\usepackage[nomath]{lmodern}  % oppure commentarlo del tutto
\usepackage{longtable} 
\usepackage{rotating}
\usepackage{array}
\usepackage{pdflscape}
\usepackage{soul} 
\usepackage{xspace}
\usepackage{enumitem}
\usepackage{hyperref}
\usepackage{color}
\usepackage{wrapfig}  %EMA
\usepackage{colortbl} %EMA
\usepackage{fix-cm}  %ALB
\usepackage{ragged2e} %ALB
\usepackage{subcaption} % ALB nov 2023

\usepackage{natbib}

\usepackage{proof}
\usepackage{makecell}

\usepackage{listings}

\newtheorem{definition}{Definition} 
\newtheorem{example}{Example} 
\newtheorem{theorem}{Theorem} 
\newtheorem{lemma}{Lemma} 
\renewcommand{\cite}{\citep}

\newcommand{\TransfTo}{\Mapsto}   
\newcommand{\Cata}{${\mathcal T}_{\mathit{cata}}$}  
\newcommand{\abs}{${\mathcal T}_{\hspace*{-.2mm}\mathit{abs}}$}    

\newenvironment{sizepar}[2]
{\par\fontsize{#1}{#2}\selectfont}
{\par}

\newcommand{\us}{\raisebox{-3pt}{\hspace{-0.2mm}-\hspace{-0.5mm}-\hspace{-0.3mm}}} % alberto 13.10.2022
 
\newcommand{\sdots}{.\hspace{.1mm}.\hspace{.1mm}.}  % alb 18.3.2024

\newcommand{\If}{\leftarrow}

\newcommand{\nots}{\sim}
\newcommand{\ands}{\scalebox{0.8}[.9]{\&}}
\newcommand{\ors}{\scalebox{0.9}[.9]{$\vee$}}

\newcommand{\newiwoADTs}{\mathit{new}1\us \mathit{woADTs}}
\newcommand{\newiiwoADTs}{\mathit{new}2\us \mathit{woADTs}}
\newcommand{\newiiiwoADTs}{\mathit{new}3\us \mathit{woADTs}}

\newcommand{\newivwoADTs}{\mathit{new}4\us \mathit{woADTs}}

\newcommand{\isasorted}{{\it is\us asorted}}

\newcommand{\vericatabs}{{VeriCaT}$_{\!\mathit{abs}}$\xspace}

\newcommand{\amps}{\scalebox{0.7}{\&}}

\begin{document}

\lefttitle{Emanuele De Angelis, Fabio Fioravanti, Alberto Pettorossi, Maurizio Proietti}

\jnlPage{1}{8}
\jnlDoiYr{2024}
\doival{10.1017/xxxxx}

\title[Catamorphic Abstractions for Constrained Horn Clause Satisfiability]
{Catamorphic Abstractions for\\ Constrained Horn Clause Satisfiability\thanks{This is an extended version of the LOPSTR 2023 paper \textit{Constrained Horn Clauses Satisfiability via Catamorphic Abstractions}, doi: \url{https://doi.org/10.1007/978-3-031-45784-5_4}}}

% \begin{authgrp}
% \author{\hspace*{-3mm}\centering{\sn{Emanuele} \gn{De Angelis} \hspace*{25.5mm} \sn{Fabio} \gn{Fioravanti}}}
% \affiliation{\hspace*{19mm}\centering{IASI-CNR, Rome, Italy \hspace*{17mm} DEc, University of Chieti-Pescara, Pescara, Italy}}
% \author{ \centering{\sn{Alberto} \gn{Pettorossi}  \hspace*{26mm} \sn{Maurizio} \gn{Proietti}}}
% \affiliation{\centering{\hspace*{-15mm}DICII, University of Rome `Tor Vergata', Rome, Italy \hspace*{12mm}IASI-CNR, Rome, Italy}}
% \end{authgrp}

\begin{authgrp}
\author{\sn{Emanuele} \gn{De Angelis}}
\affiliation{IASI-CNR, Rome, Italy}
\author{\sn{Fabio} \gn{Fioravanti}}
\affiliation{DEc, University of Chieti-Pescara, Pescara, Italy}	
\author{\sn{Alberto} \gn{Pettorossi}}
\affiliation{DICII, University of Rome `Tor Vergata', Rome, Italy}
\author{\sn{Maurizio} \gn{Proietti}}
\affiliation{IASI-CNR, Rome, Italy}
\end{authgrp}

%%\begin{authgrp}
%%\author{\sn{Emanuele} \gn{De Angelis}}
%%\affiliation{IASI-CNR, Rome, Italy}
%%\author{\sn{Fabio} \gn{Fioravanti}}
%%\affiliation{DEc, University `G.\,d'Annunzio', Chieti-Pescara, Italy}
%%\author{\sn{Alberto} \gn{Pettorossi}}
%%\affiliation{DICII, University of Rome `Tor Vergata', Rome, Italy}
%%\author{\sn{Maurizio} \gn{Proietti}}
%%\affiliation{IASI-CNR, Rome, Italy}
%%\end{authgrp}

%%%%%\authorrunning{E.~De~Angelis, F.~Fioravanti, A.~Pettorossi, and M.~Proietti}
%%%%%
%%%%%\institute{IASI-CNR, Rome, Italy
%%%%%	\email{\{emanuele.deangelis,maurizio.proietti\}@iasi.cnr.it}
%%%%%	\and DEc,\,University\,`G.\,d'Annunzio',\,Chieti-Pescara,\,Italy
%%%%%	\email{fabio.fioravanti@unich.it}
%%%%%	\and DICII, University of Rome `Tor Vergata', Italy
%%%%%	\email{pettorossi@info.uniroma2.it}
%%%%%}

%\history{\sub{xx xx xxxx;} \rev{xx xx xxxx;} \acc{xx xx xxxx}}

\maketitle

\begin{abstract}
Catamorphisms are functions that are recursively defined on list and trees and, in general,
on Algebraic Data Types (ADTs),
%%Catamorphisms are functions that are recursively defined on Algebraic Data Types 
%%(ADTs) such as lists or trees. 
and are often used to compute suitable abstractions
of programs that manipulate ADTs. 
Examples of catamorphisms include functions that compute %, for instance, 
size of lists, orderedness of lists, and height of trees.
It is well known that program properties specified through catamorphisms can be 
proved by showing the
satisfiability of suitable sets of Constrained Horn Clauses (CHCs).
We address the problem of checking the satisfiability of 
those sets of CHCs, % that encode program properties,
%that kind of program properties, 
and we propose a method for transforming sets %a given set 
of CHCs %satisfiability problem 
into equisatisfiable sets where catamorphisms are no longer present. 
As a consequence, clauses with catamorphisms 
can be handled without extending the satisfiability algorithms 
used by %present in  
existing CHC solvers. Through an experimental evaluation on a 
non-trivial benchmark consisting of many list and tree processing
%programs, 
algorithms expressed as sets of CHCs, we show that our technique is indeed effective and significantly enhances the performance of
state-of-the-art CHC solvers.

\medskip
\noindent
\textit{Under consideration in Theory and Practice of Logic Programming (TPLP)}

\end{abstract}


\section{Introduction} 
\label{sec:Intro}
%\vspace*{-2mm}
Catamorphisms are functions that compute abstractions over Algebraic Data Types  (ADTs), such as lists or trees.
The definition of a catamorphism is based on a simple recursion scheme, called a \emph{fold} in the context of functional programming~\cite{MeijerFP91}. Examples of  catamorphisms on lists of integers include functions that compute the orderedness of a list, the length of a list, and the sum of its elements. Similarly, examples of catamorphisms on trees are functions that compute the size of a tree, the height of a tree, and the minimum integer value at its nodes.

Through catamorphisms we can specify many useful program properties such as, for instance, the property that the list computed by a program for sorting lists is
indeed sorted, or the property that the output list has the same length of the input list. For this reason, program analysis tools based on \emph{abstract interpretation}~\cite{CoC77,HermenegildoPBL05} and \emph{program verifiers}~\cite{Su&11} have implemented special purpose techniques that handle catamorphisms.

In recent years, it has been shown that verification problems that 
%are based on the use of 
use catamorphisms can be reduced to satisfiability problems for 
Constrained Horn Clauses (CHCs) by following a general approach that 
is very well suited for automatic 
proofs~\cite{Bj&15,DeAngelisFGHPP21,Gurfinkel22}.
%program verification~\cite{Bj&15,DeAngelisFGHPP21,Gurfinkel22}. 
A practical advantage of CHC-based verification is that it is 
supported by several %effective 
CHC \textit{solvers} which can be used as back-end 
tools~\cite{BlichaFHS22,chccomp22,HoR18,Ko&16}. 

Unfortunately, the direct translation of catamorphism-based 
verification problems into CHCs is not always helpful, because CHC 
solvers often lack %may lack 
mechanisms for computing solutions %models 
by performing 
induction over ADTs. To overcome this difficulty, some CHC solvers 
have been extended 
with special purpose satisfiability algorithms that handle 
(some classes of) 
catamorphisms~\cite{GovindSG22,HoR18,KostyukovMF21}. For instance, the module of Eldarica
for solving CHCs has been extended by allowing constraints that use the
built-in \textit{size} function counting the number of function symbols in the 
ADTs~\cite{HoR18}. 

\sloppy
In this paper {we consider a class of catamorphisms that is strictly larger than the ones handled by the above mentioned satisfiability algorithms} and we follow an approach based on the transformation of CHCs~\cite{DeAngelisFPP22,DeAngelisFPP23a}.
% of sets of CHCs.
In particular, given a set $P$ of CHCs that uses catamorphisms and includes one or 
more queries encoding the properties of 
interest, we transform $P$ into a new set $P'$ such 
that: (i) $P$ is 
satisfiable if and only if $P'$ is satisfiable, and (ii) no 
catamorphism is present in $P'$. 
%%\comment{MAU: not true in our running example, \textit{listcount} appears in the final program. {\color{blue}{alb: come risolviamo questa difficolta'? Il sistema produce new4 ed elimina listcount: vero? MAU: corretto.}}}
%%%%
%%Teniamo presente che nella risposta ai referre diciamo:
%%"they can be handled without any ad-hoc satisfiability algorithm". Se necessario, dobbiamo modificare anche la risposta ai referee.
%%MAU: questa frase rimane corretta.
%%
%%Tecnicamente, vengono eliminate le congiunzioni p(L,....), cata(L,...) in favore di newp(L,...), newp\_w0(...) }. Difficile da spiegare in maniera non criptica.
%%
%%Forse conviene seguire l'implementazione nell'esempio e rimpiazzare listcount con new4 e new4\_wo. (risparmieremmo anche le frasi "together with clauses 5-6 for listcount".)}}
Thus, the satisfiability of $P'$ can be verified by a CHC solver that 
is not extended 
for handling catamorphisms. % and performing induction over~ADTs.

\fussy

The main difference between the technique we present in this paper and 
the above cited 
works~\cite{DeAngelisFPP22,DeAngelisFPP23a} is that the algorithm we present here does 
not require that we specify suitable properties of how the 
catamorphisms relate to every predicate occurring 
in the given set $P$ of CHCs.
%%, that we specify suitable properties of the 
%%catamorphisms when they act on that predicate. %occurring in the given set $P$ of CHCs. 
%%not require, for each predicate in the given set $P$ of CHCs, a specification of how the 
%%catamorphisms relate to that predicate. 
For instance, if we want to verify
that the output list $S$ of the set of CHCs defining ${\textit{quicksort}}(L,S)$ has the same length of the 
input list $L$, we need not specify that, for the auxiliary predicate 
${\textit{partition}}(X,\textit{Xs},\textit{Ys},\textit{Zs})$ that divides  the list $\textit{Xs}$ into the two 
lists $\textit{Ys}$ and $\textit{Zs}$, it is the case that the length of~$\textit{Xs}$ is the
sum of the lengths of $\textit{Ys}$ and $\textit{Zs}$. This property {can} %will 
automatically be derived %discovered 
by the CHC solver when it looks for a model of the set of CHCs 
obtained by transformation. 
%\comment{and looks for a model of those clauses.}
{In this sense, our technique may allow the discovery of some %the suitable 
lemmas needed for the proof of the property of interest.}
%Similarly, it is not required to state how the lengths of the lists 
%$\mathit{Xs},\!\mathit{Ys},$ and $\mathit{Zs}$ are related for the predicate \textit{append}$(\mathit{Xs},\!\mathit{Ys},\!\mathit{Zs})$ 
%which also occurs  in the program ${\textit{quicksort}}(L,S)$.
%Thus, the technique we propose in this paper for  property verification
% turns out to be significantly more useful  with respect to the 
% techniques that have been proposed in previous papers.

We will show through a benchmark set of list and tree processing algorithms
expressed as sets of CHCs,
that our 
transformation technique is indeed effective and is able to 
drastically increase the 
performance of state-of-the-art CHC solvers such as 
Eldarica~\cite{HoR18} 
(with the built-in catamorphism \textit{size}) and Z3 with the SPACER  engine~\cite{DeB08,Ko&16}.

The rest of the paper is organized as follows.
In Section~\ref{sec:CHCs} we recall some preliminary notions on CHCs and catamorphisms.
In Section~\ref{sec:IntroExample} we show an introductory example to 
motivate our technique. 
%\comment{+++ review section numbers}
In Section~\ref{sec:Cata} we present our transformation algorithm and
 prove that it guarantees the equisatisfiability of the initial sets of CHCs and the transformed sets of CHCs. 
In Section~\ref{sec:Experiments} we present the implementation of our technique in the \vericatabs tool, and  through an experimental evaluation, we show the beneficial effect of the transformation on both Eldarica and Z3 CHC solvers. We will consider several abstractions based on catamorphisms relative to lists and trees, such as size, minimum element, orderedness, element membership, element multiplicity, and combinations thereof. 
%%\comment{Those abstractions will be called  
%%\textit{catamorphic abstractions}.} 
Finally, in Section~\ref{sec:RelConcl}, we discuss related work and we outline future research directions.

\section{Basic Notions} 
\label{sec:CHCs}
%\comment{from ICLP 2022. Change the presentation a little bit to avoid plagiarism.}
%\newcommand{\cd}{:\!\scalebox{1.7}[1.2]{-}~} % simulating in math the :- symbol of Prolog

%\comment{ NO dots at the end of clauses and abstraction specs}

The programs and the properties we consider in this paper are expressed
as sets of  constrained Horn clauses written in a
many-sorted first-order language $\cal L$ with equality (=).
Constraints are expressions of the linear integer arithmetic (\textit{LIA})
and the boolean algebra~(\textit{Bool\/}).
The theories of \textit{LIA} and \textit{Bool\/} will be collectively
denoted by $\mathit{LIA\cup Bool}$. The equality symbol = will be used both for integers and booleans.
In particular, a {\em constraint} is a quantifier-free formula $c$, where
\textit{LIA} constraints may occur as subexpressions of
boolean constraints, according to the SMT approach~\cite{Ba&09}. The
syntaxes of a constraint $c$ and an elementary \textit{LIA}  constraint $d$ are as follows:

\vspace{1.mm}
\hspace{-3mm}
$c~::=~d~~|~~\textit{B}~~|~~\textit{true}~~|~~\textit{false}~~|~\nots\!c~~|~~c_1 \,\ands\,c_2~~|~~c_1\ors\, c_2~~|~~c_1\!\Rightarrow\!c_2~|~~c_1\!=\!c_2~~|$~~

%\vspace{.5mm}

\hspace*{8mm}$\textit{ite}(c, c_1, c_2)~~|~~t\!=\!\textit{ite}(c, t_1, t_2)$
\vspace{1.mm}

\hspace{-3mm}
$d~::=~t_1\!\!<\!t_2~~|~~t_1\!\!\leq\!t_2~~|~~t_1\!\!=\!t_2~~|~~t_1\!\geq t_2~~|~~t_1\!\!>\!t_2 $

\vspace{1.mm}
\noindent
where:
%$r$ is a \textit{LIA} atomic formula of the form
%$r ::=\, t_1\!\!=\!t_2 \,|\, t_1\!\!\geq\!t_2 \,|\, t_1\!\!>\!t_2 \,|\,
%t_1\!\!\leq\!t_2 \,|\, t_1\!\!<\!t_2$;
(i)~{$B$} is a boolean variable, (ii)~$\nots,\ \ands$, \ors, and $\Rightarrow$
denote negation, conjunction, disjunction, and implication, respectively, (iii)~the ternary
function \textit{ite} denotes the % usual 
\mbox{if-then-else} operator
(that is, 
$\mathit{ite}(c, c_1, c_2)$ has the following semantics: if  $c$  then $c_{1}$  else $c_{2}$),
% 
%$\mathit{ite}(c, c_1, c_2)$ holds iff $c$ and $c_{1}$ hold or +++ e
%
%
%the semantics of $\mathit{ite}(c, c_1, c_2)$ is the following: 
%$\textrm{if}\ c\  \textrm{holds\ in}\  \mathit{LIA\cup Bool}$ $\textrm{then}\ c_{1}$ 
%$\textrm{else}\ c_{2}$)}, 
and (iv)~{$t$}, possibly with subscripts,
$t$, $t_{1}$ and $t_{2}$
is a \textit{LIA}~term of the form {$a_0+a_1X_1+\sdots+a_nX_n$}
with integer coefficients {$a_0,\dots,a_n$}
and integer variables {$X_1,\sdots,X_n$}.
% We will often write the constraints of the form
% $B1\!=\!\mathit{true}$ and
% $B2\!=\!\mathit{false}$ as $B1$ and $\nots\!\!B2$, respectively.
% The theory of \textit{LIA} and boolean constraints will be denoted by
% $\textit{LIA}\cup\textit{Bool\/}$.

%%  Not inserted: ALB 2024-01-24
%%We will allow ourselves to write the constraint \mbox{$B\!=\!\mathit{true}$} as $B$, and 
%%the constraint \mbox{$B\!=\!\mathit{false}$} as \mbox{$\nots\!\!B$}.

The integer and boolean sorts are said to be %the
{\it basic sorts}. 
A recursively defined sort (such as the sort of lists and trees) 
is said to be an {\it algebraic data type} (ADT, for short).

An {\it atom} is a formula of the form $p(t_{1},\ldots,t_{m})$, where~$p$
is a predicate symbol not occurring in $\textit{LIA}\cup\textit{Bool\/}$,
and $t_{1},\ldots,t_{m}$ are first-order terms in $\cal L$.
A~{\it constrained Horn clause}  (CHC), or simply, a {\it clause}, is
an implication of the form $H\leftarrow c, G$.
The conclusion~$H$, called the {\it head\/}, is either an atom or \textit{false}, and
the premise, called the {\it body\/}, is the conjunction of a constraint~$c$
and a conjunction~$G$ of zero or more atoms. $G$ is said to be a {\it{goal\/}}.
A clause is said to be a {\it  query\/} if its head is \textit{false}, 
and
a {\it definite clause\/}, otherwise.
Without loss of generality, at the expense of introducing suitable equalities, we assume that every atom of the body
of a clause has distinct variables (of
any sort) as arguments. 
%\bcomment{(but two atoms may share a variable)}.
%
Given an {expression} $e$, by ${\it vars}(e)$ we denote the set of 
all variables occurring in~$e$.
By $\mathit{bvars(e)}$ (or $\mathit{adt}$-$\mathit{vars}(e)$) we
denote the set of variables in~$e$ whose sort is a basic sort 
(or an ADT sort, respectively).
The {\it universal closure} of a formula~$\varphi$ is denoted by 
$\forall (\varphi)$.

%Let~$\mathbb D$ be the usual interpretation for the symbols of {theory}
%$\textit{LIA}\cup\textit{Bool\/}$. 
A $\mathbb D$-interpretation for a set $S$ of CHCs is an interpretation where the symbols of $\textit{LIA}\cup\textit{Bool\/}$ are interpreted as usual. 
A  $\mathbb D$-interpretation $I$ is said to be a $\mathbb D$-model of $S$ if all clauses of $S$ are true in $I$.
A set $S$ of CHCs is said to be ${\mathbb D}$-{\it satisfiable} (or {\it satisfiable}, for short) 
if it has a ${\mathbb D}$-model, and it is said to be ${\mathbb D}$-{\it unsatisfiable} (or {\it unsatisfiable}, for short), otherwise.

Given a set $P$ of definite clauses, there exists a  {\it least} ${\mathbb D}$-model of $P$, 
denoted $M(P)$ \cite{JaM94}.
Let $P$ be a set of definite clauses and for $i=1,\ldots,n$, $Q_i$  be a  query. Then
$P \cup \{Q_1,\ldots,Q_n\}$ is satisfiable if and only if, for $i=1,\dots,n$, $M(P)\models Q_i$.

\smallskip

%Now, in order to characterize the class of queries that can be handled using our transformation
%technique, we introduce (see Definition~\ref{fig:cataExt} below) a class of recursive schemata defined by CHCs~\cite{DeAngelisFPP22}.
%Similar classes have been considered in the field of functional programming where
%various kinds of recursive functions, called \textit{morphisms}, have been introduced and
%turned out to be very useful~\cite{HinzeWG13,MeijerFP91}.
%%
%With reference to those
%recursive functions, our CHC schemata can be viewed, in the logic programming setting,
%as a generalization of the \textit{catamorphisms} and
%a specific form of the \textit{paramorphisms}~\cite{MeijerFP91}. For reasons
%of simplicity, we will not introduce a new terminology
%%

The catamorphisms we consider in this paper are defined by first-order, relational recursive schemata as we now indicate. Similar definitions are introduced also in (higher-order) functional programming~\cite{HinzeWG13,MeijerFP91}. 

Let $f$ be a predicate symbol with $m + n$ arguments (for $m\!\geq\! 0$ and $n\!\geq\! 0$) with sorts
\! $\alpha_1,\ldots,\alpha_m,$ $\beta_1,\ldots,\beta_n$, respectively.
%Let $f$ be a predicate symbol of arity $m\plus n$, with $m,n\geq 0$, whose arguments
%have sorts $\alpha_1,\ldots,\alpha_m,$ $\beta_1,\ldots,\beta_n$, respectively.
We say that $f$ is a {\em functional predicate} from sort 
$\alpha_1\!\times\!\ldots\!\times\!\alpha_m$ to  sort $\beta_1\!\times\!\ldots\!\times\!\beta_n$,
with respect to a given set $P$ of definite clauses that define $f$,
if $M(P)\! \models\! \forall X,\!Y,\! Z.\ f(X,\!Y) \wedge f(X,\!Z) \rightarrow Y\!=\!Z$,
where $X$ is an $m$-tuple of distinct variables, and
$Y$ and $Z$ are $n$-tuples of distinct variables.
In this case, when we write the atom $f(X,Y)$,
we mean that 
$X$ and~$Y$ are the tuples of the {\em input}
and {\em output} variables of $f$, respectively.
%Given the atom $f(X,Y)$, we say that
%$X$ and $Y$ are the tuples of the {\em input}
%and {\em output} variables of $f$, respectively.
%
We say that~$f$ is a {\em total predicate} if $M(P) \models \forall X \exists Y.\ f(X,Y)$.
In what follows, a `total, functional predicate' $f$ from 
a tuple~$\alpha$ of sorts to a tuple~$\beta$ of sorts
is said to be % will be called 
a `total function' in $\mbox{[}\alpha \rightarrow \beta\mbox{]}$, 
%\comment{(or simply, a {\it function})}
and it is denoted by $f \in \mbox{[}\alpha \rightarrow \beta\mbox{]}$.

Now we introduce the notions of a list catamorphism and 
 a binary tree catamorphism. 
We leave to the reader the task of introducing, the definitions of 
similar catamorphisms 
for recursively defined algebraic data types that may be needed for
expressing the properties of interest.
%%(the set $P$ of clauses that define $f$ will be understood 
%% from the context).
Let~$\alpha$, $\beta$,
$\gamma$, and $\delta$ be $($products of$\,)$ basic sorts. Let 
$\mathit{list}(\beta)$ be the sort of lists with elements of sort $\beta$, 
and $\mathit{btree}(\beta)$ be the sort of binary trees with values of sort $\beta$.

\vspace*{-1mm}
\begin{definition}[List and Binary Tree Catamorphisms]\label{def:listbtree-cata}
A {\em list catamorphism} $\ell$ %$($see Figure~{\rm\ref{fig:listcata}}$)$ 
is a total function in $[\alpha\!\times\!\mathit{list}(\beta)
\rightarrow \gamma]$ defined as follows\,$:$

\vspace{1mm}
\begin{sizepar}{9}{11}
$L1.~~~\ell(X,[\,],Y) \If   \mathit{\ell\us basis}(X,Y)$

$L2.~~~\ell(X,[H|T],Y) \If f(X,T,\mathit{Rf}),\  \ell(X,T,R),\  
\mathit{\ell\us combine}(X,H,R,\mathit{Rf},Y)$
\end{sizepar}

\vspace{1mm}
\noindent
where$\,:$ $({\rm{i}})~\mathit{\ell\us basis}$ $\in$ $[\alpha\!\rightarrow\!\gamma]$,~
$({\rm{ii}})~\mathit{\ell\us combine}$ $\in$
$[\alpha\!\times\!\beta\!\times\!\gamma\!\times\!\delta\rightarrow \gamma]$,
and $({\rm{iii}})~f$ is itself a list catamorphism in 
\mbox{$[\alpha\!\times\!\mathit{list}(\beta) \rightarrow \delta]$}.

\noindent
A {\em binary tree catamorphism}  $\mathit{bt}$ is a
total function in $[\alpha\!\times\!\mathit{btree}(\beta)
\rightarrow \gamma]$ defined as follows\,$:$~
\vspace{1mm}
\begin{sizepar}{9}{11}

$\mathit{BT}1.~~\mathit{bt}(X,\mathit{leaf},Y) \If  \mathit{bt\us basis}(X,Y)$

$\mathit{BT}2.~~\mathit{bt}(X,\mathit{node}(L,N,R),Y) \If  
g(X,L,\mathit{RLg}),\  g(X,R,\mathit{RRg}),$

\hspace*{15mm}$\mathit{bt}(X,L,\mathit{RL}),\  \mathit{bt}(X,R,\mathit{RR}),\  
\mathit{bt\us combine}(X,N,\mathit{RL,RR,RLg,RRg},Y)$
\end{sizepar}
\vspace{1mm}

\noindent
where$\,:$ $({\rm{i}})~\mathit{bt\us basis}$ $\in$ $[\alpha\rightarrow \gamma]$,
$({\rm{ii}})$~$\mathit{bt\us combine}$  $\in$
$[\alpha\!\times\!\beta\!\times\!\gamma\!\times\!\gamma\!\times\!\delta\!\times\!\delta\rightarrow \gamma]$, and
$({\rm{iii}})~g$ is itself a binary tree catamorphism in 
$[\alpha\!\times\!\mathit{btree}(\beta)
\rightarrow \delta]$.
\end{definition}

\vspace*{-1mm}
%Examples of catamorphisms will be shown in the following sections. In particular, 

%%Instances of the schemas of the catamorphisms $\ell$ and 
%%$\mathit{bt}$ of Definition~\ref{def:listbtree-cata} 
Instances of the schemas of the list catamorphisms and the binary tree
catamorphisms (see  
Definition~\ref{def:listbtree-cata} above) 
may lack some components, such as the parameter~$X$ of basic sort~$\alpha$, or the catamorphisms $f$ or $g$. 
%or the catamorphism $f$. 
{The possible presence of these components 
makes the class of catamorphisms considered in this paper strictly larger than the ones used by other CHC-based approaches~\cite{GovindSG22,HoR18,KostyukovMF21}.}

\vspace*{-3mm}

\section{An Introductory Example} 
\label{sec:IntroExample}
%

%Figure~\ref{TransfProgDefs}

Let us consider %an introductory example consisting of 
a set of CHCs for doubling lists of integers (see 
clauses 1--4 in Figure~\ref{fig:ProgCataQuery}). 
We have that: (i)~$\mathit{double}(\mathit{Xs},\mathit{Zs})$ holds if and only if  
list~\textit{Zs} is the concatenation
of two copies of the same list \textit{Xs} of integers, (ii)~$\mathit{eq(Xs,\!Ys)}$ holds if and only if 
list~$\mathit{Xs}$ is equal to list $\mathit{Ys}$, and 
(iii)~$\mathit{append(Xs,Ys,Zs)}$ holds if and only if  list~\textit{Zs} is the result 
of concatenating
list~\textit{Ys} to the right of list~\textit{Xs}.
 
Let us assume that we want to verify 
the following {\it Even} property: %stating that
% the following property of the atom 
if $\mathit{double}(\mathit{Xs},\mathit{Zs})$
holds, then
for any integer~$X$, the number of occurrences of $X$ in \textit{Zs} is an 
even number. 
In order to do so, we use the list catamorphism 
$\mathit{listcount(X,Zs,M)}$ (see clauses~5--6 
in Figure~\ref{fig:ProgCataQuery}) that holds if and only if 
$M$ is the number of occurrences of $X$ in list~$\mathit{Zs}$. 
Note that $\mathit{listcount(X,Zs,M)}$ is indeed a list catamorphism because clauses 5--6 
%defining the catamorphism $\mathit{count}$,
are instances of clauses $L1$--$L2$ in Definition~\ref{def:listbtree-cata}, when: (i)\,$\ell$ is \textit{listcount},
(ii)\,$Y$ is~$N$, (iii)\,$\mathit{\ell\us basis}(X,Y)$ is the {\it LIA} constraint $N\!=\!0$, (iv)\,$f(X,T,\mathit{Rf})$ is absent, and (v)\,$\mathit{\ell\us combine}(X,H,R,\mathit{Rf},Y)$ is the {\it LIA} %if-then-else 
constraint $N\!=\mathit{ite}(X\!\!=\!\!H, \mathit{NT}\!+\!{\rm 1},\mathit{NT})$.

Our verification 
task can be expressed as query 7 in Figure~\ref{fig:ProgCataQuery}, whereby we derive 
$\mathit{false}$ if the number $M$ of occurrences of~$X$ in $\mathit{Zs}$ is odd (recall that we assume that
$M \!=\! 2\,N\!+\!1$ is a $\mathit{LIA}$ constraint).  
\begin{figure}[!ht]
\hspace*{-3mm}\hrulefill
\vspace*{-2mm}
\begin{sizepar}{9.3}{11}
\begin{justify}
*** Initial set of CHCs including the catamorphism $\mathit{listcount}$.
\vspace{1mm}

~1. $\mathit{double(Xs,\!Zs) \leftarrow eq(Xs,\!Ys),\ append(Xs,\!Ys,\!Zs)}$

~2. $\mathit{eq(Xs,\!Xs)} \leftarrow$

~3. $\mathit{append([\,],\!Ys,\!Ys)}\leftarrow$ 

~4. $\mathit{append([X|Xs],\!Ys,\![X|Zs]) \leftarrow append(Xs,\!Ys,\!Zs)}$

~5. $\mathit{listcount(X,[\,],N) \leftarrow N\!=\!{\rm 0}}$

~6. $\mathit{listcount(X,[H|T],N) \leftarrow N\!=\!\mathit{ite}(X\!\!=\!\!H, NT\!+\!{\rm 1},NT),\ listcount(X,T,NT)}$

\vspace{2mm}
\noindent
*** Query.

~7. $\mathit{false} \leftarrow M \!=\! 2\,N\!+\!1,\ \mathit{listcount}(X,\mathit{Zs},M),\ \mathit{double}(\mathit{Xs},\mathit{Zs})$
  
\vspace*{-2mm}
\end{justify}
\end{sizepar}

\vspace*{-1mm}
\hspace*{-3mm}\hrulefill
\vspace*{-1mm}
%%\caption{The initial CHCs (1--4), the catamorphism $\mathit{count}$
%%(5--6),  and query~7 specifying the property of interest. 
%and the abstractions introduced ($A1$--$A3$). 
\caption{The initial set of CHCs (clauses 1--6) and query~7 that specifies that the number of occurrences of an element
$X$ in the list $\mathit{Zs}$ is even. %the property of interest. 
%%Clauses 5--6 defining the catamorphism $\mathit{count}$,
%%are instances of the list catamorphism clauses of Definition~\ref{def:listbtree-cata}, where: (i)\,predicate $\ell$ is \textit{count}, (ii)\,predicate  $\mathit{\ell\us basis}$ is the constraint $ N\!=\!{\rm 0}$, (iii)~predicate\,$f$ is absent, and (iv)\,predicate $\mathit{\ell\us combine}$ is the %if-then-else 
%%constraint $N\!=\mathit{ite}(X\!\!=\!\!H, \mathit{NT}\!+\!{\rm 1},\mathit{NT})$.
\label{fig:ProgCataQuery}}
\vspace*{-1mm}
\end{figure}

Now, neither the CHC solver Eldarica nor Z3 %(with the SPACER engine) 
is able to prove the satisfiability of clauses~1--7 and thus, 
those solvers are not able to show the {\it Even} property.
By the transformation technique we will propose in this paper, we get a new set of clauses %(see Figure~\ref{fig:TransfProgDefs}) 
whose satisfiability can be shown by Z3 and thus, the {\it Even} property is proved.

To perform this transformation, we use the information that the property to be verified is expressed through the catamorphism $\mathit{listcount}$.
However, in contrast to previous approaches 
%described in the literature
\cite{DeAngelisFPP22,DeAngelisFPP23a}, we 
need {\it not\/} specify 
any property of the catamorphism $\mathit{listcount}$ when it acts upon
the predicates $\mathit{double}$, $\mathit{eq}$, and $\mathit{append}$. 
For instance, we need not specify that if \textit{Zs} is the concatenation of \textit{Xs} and \textit{Ys}, then
for any $X$, the number of occurrences of $X$ in \textit{Zs} is the sum of the numbers of occurrences 
of $X$ in $\mathit{Xs}$ and~$\mathit{Ys}$.
Indeed, in the approach we propose in this paper, %according the %novel 
%technique we propose in this paper, before starting the clause transformation process,
we have only to specify the 
association of every ADT sort with a suitable catamorphism 
or, in general, %possibly empty 
a conjunction of catamorphisms. 
%%(or, in general, %possibly empty 
%%a finite, possibly empty, conjunction of distinct catamorphisms). 
In particular, in our {\it double} example, we associate
the sort of integer lists, denoted 
$\mathit{list(int)}$, with the catamorphism $\mathit{listcount}$.
%By applying this technique, very little ingenuity is required on the part of the programmer 
%for specifying and verifying program correctness.
Then, we rely on the CHC solver for the discovery, after the transformation described in the following sections, of suitable relations between the variables that represent the output of the %some 
\textit{listcount} catamorphism atoms. Thus, by applying the technique proposed in this paper, much less ingenuity is required on the part of the programmer 
for verifying program correctness with respect to the 
previously proposed approaches.

%Now, neither the CHC solver Eldarica nor Z3 with the SPACER engine is able to prove the satisfiability of the initial set of CHCs (clauses 1--7) shown in Figure \ref{fig:ProgCataQuery} and thus, 
%those solvers are not able to show the {\it Even} property.
%% of $\mathit{double}(\mathit{Xs},\mathit{Zs})$.
%However, if we transform those initial clauses according to our novel technique, % we propose in this paper, 
%we get a set of clauses (see Figure~\ref{fig:TransfProgDefs}) whose satisfiability can be shown by Z3 and thus, the {\it Even} property is proved.

%In particular, in the case of our {\it double} example,
%sort  $\mathit{list}({\mathit{int}})$

Our transformation technique introduces,  %our novel technique %we propose in this paper 
%starts off by introducing,
for each predicate~$p$ occurring in the initial set of CHCs, a new predicate 
$\textit{newp}$ defined by the conjunction of a~$p$ atom and, for each argument of~$p$ 
with ADT sort~$\tau$, the catamorphism atom(s) with which $\tau$ has been associated.
%a catamorphism atom associated with $\tau$.
In particular, in the case of our {\it double} example,
for the predicate $\mathit{double}$ 
%for the atom $\mathit{double}(B,E)$ 
we introduce the new predicate $\mathit{new}1$ (for 
simplicity, we call it $\mathit{new}1$, instead of $\mathit{newdouble}$) whose 
definition is clause~$D1$ in Figure~\ref{fig:TransfProgDefs}. The body of that clause
is the conjunction of the atom $\mathit{double}(B,E)$ and two {\it listcount} catamorphism atoms, one for each of the integer lists $B$ and $E$, as {\it listcount} is the catamorphism with which the sort of integer lists has been associated. 
%%one for the integer list $B$ and one 
%%for the integer list $E$, as {\it count} is the catamorphism 
%%with which the sort of integer lists has been associated. 
%which is associated 
%with the sort of integer lists. 
% $\mathit{list(int)}$.
%Similarly, for the atoms $\mathit{append(H,B,E)}$ and $\mathit{eq}(E,B)$ for 
Similarly, for the predicates $\mathit{append}$ and $\mathit{eq}$  
whose 
%for which we introduce the  
definitions are respectively clauses~$D2$ and $D3$ listed in Figure~\ref{fig:TransfProgDefs}.

%%
%%every definition introduced 
%%%sort  $\mathit{list}({\mathit{int}})$Note that the new predicate definitions introduced 
%%during the transformation process (see clauses $D1$--$D3$ of 
%%Figure~\ref{fig:TransfProgDefs}) has in its body an atom and for each variable of sort  
%%$\mathit{list}({\mathit{int}})$ in that atom, it has an associated 
%%catamorphism \textit{count}.
%%For instance, the body of definition $D1$ 
%%is the conjunction of the
%%$\mathit{double}(\underline{B},\!\underline{E})$ 
%%atom and two catamorphisms $\mathit{count}$, one for the integer list 
%%$\underline{B}$ and one for the integer list $\underline{E}$.
%%For instance, the body of definition $D2$ 
%%is the conjunction of 
%%$\mathit{append}(\underline{H},\!\underline{B},\!\underline{E})$ 
%%and three catamorphisms $\mathit{count}$, one for each of the integer lists 
%%$\underline{H}, \underline{B},$ and $\underline{E}$.

\begin{figure}[!ht]
\hrule\vspace{2mm}
\begin{sizepar}{9.3}{11}
\begin{justify}
\hangindent=7mm
*** The new predicate definitions introduced during transformation. 
%In these definitions, 
%For the convenience of the reader, 
%The underlined variables are those with an ADT sort, which is the sort of the integer lists. 
In these definitions the underlined variables have an ADT sort (i.e., they are integer lists),
while the non-underlined variables have a basic sort (i.e., they are integer numbers). %  have basic sort of integers.

%%the ADT variables. Their sort is $\mathit{list}(\mathit{int})$.
%%The non-underlined variables have integer sort. 

\vspace{1mm}
%\begin{sizepar}{8.5}{11}

$D1$. $\mathit{new}1(A,\underline{B},C,\underline{E},F) \leftarrow \mathit{listcount}(A,\underline{B},C),\  \mathit{listcount}(A,\underline{E},F),\ \mathit{double}(\underline{B},\underline{E})$

$D2$. $\mathit{new}2(\!A,\!\underline{B},\!C,\!\underline{E},\!F,\!\underline{H},\!I) \leftarrow \mathit{listcount}(\!A,\!\underline{B},\!C),\  \mathit{listcount}(\!A,\!\underline{E},\!F),\  \mathit{listcount}(\!A,\!\underline{H},\!I),$ % \mathit{append}(\underline{H},\!\underline{B},\!\underline{E})$

\hspace*{45mm}$\mathit{append}(\underline{H},\!\underline{B},\!\underline{E})$

$D3$. $\mathit{new}3(A,\underline{B},C,\underline{E},F) \leftarrow \mathit{listcount}(A,\underline{B},C),\  \mathit{listcount}(A,\underline{E},F),\  \mathit{eq}(\underline{E},\underline{B})$

$D4$. $\mathit{new}4(A,\underline{B},C) \leftarrow \mathit{listcount}(A,\underline{B},C)$

\vspace{3mm}
\noindent
*** Clauses derived after transformation from clauses 1--7 of Figure~\ref{fig:ProgCataQuery}.
\vspace{1mm}

11. $\mathit{false} \leftarrow C\!=\!2\,D\!+\!1,\ \mathit{new}1(A,E,F,G,C),\ \newiwoADTs(A,F,C)$

12. $\mathit{new}1(A,B,C,E,F) \leftarrow \mathit{new}2(A,M,K,E,F,B,C),\ \newiiwoADTs(A,K,F,C),\ $
      
      \hspace{20mm}
      $\mathit{new}3(A,M,K,B,C),\ \newiiiwoADTs(A,K,C)$
      
13. $\mathit{new}2(A,\!B,\!C,B,C,[\,],\!G) \leftarrow G\!=\!0,\ \mathit{new}4(A,B,C),\ \newivwoADTs(A,C)$

% Here below three times P replaced O (capital O) to avoid confusion 
% with zero.
14. $\mathit{new}2(A,\!B,\!C,[E|F],G,[E|J],\!K)\! \leftarrow\! 
      G\!=\!\mathit{ite}(A\!=\!E,\!N\!+\!1,\!N),\ K\!=\!\mathit{ite}(A\!=\!E,\!P\!+\!1,\!P),\ $
      
      \hspace{20mm}
      $\mathit{new}2(A,B,C,F,N,J,P),\ \newiiwoADTs(A,C,N,P)$
      
15. $\mathit{new}3(A,B,C,B,C) \leftarrow \mathit{new}4(A,B,C),\ \newivwoADTs(A,C)$

16. $\mathit{new}4(A,[\,],B) \leftarrow  B\!=\!0$

17. $\mathit{new}4(A,[B|C],D) \leftarrow  D\!=\!\mathit{ite}(A\!=\!B,F\!+\!1,F),\ \mathit{new}4(A,C,F),\ \newivwoADTs(A,F)$\\

18. $\newiwoADTs(A,B,D) \leftarrow \newiiwoADTs(A,I,D,B),\ \newiiiwoADTs(A,I,B)$

19. $\newiiwoADTs(A,B,B,F) \leftarrow F\!=\!0,\ \newivwoADTs(A,B)$

20. $\newiiwoADTs(A,B,D,F) \leftarrow 
  D\!=\!\mathit{ite}(A\!=\!I,K\!+\!1,K),\ F\!=\!\mathit{ite}(A\!=\!I,L\!+\!1,L),\ $
       
      \hspace{20mm}$\newiiwoADTs(A,B,K,L)$

21. $\newiiiwoADTs(A,B,B) \leftarrow \newivwoADTs(A,B)$
      
%%20. $\mathit{\listcountwoADTs}(A,B) \leftarrow B\!=\!0$
%%
%%21. $\mathit{\listcountwoADTs}(A,B) \leftarrow B\!=\!\mathit{ite}(A\!=\!C,D\!+\!1,D),\  
%%\mathit{\listcountwoADTs}(A,D)$

22. $\newivwoADTs(A,B) \leftarrow B\!=\!0.$

23. $\newivwoADTs(A,B) \leftarrow B\!=\!\mathit{ite}(A\!=\!C,D\!+\!1,D),\ \newivwoADTs(A,D).$

\vspace*{-1mm}
\hspace*{-3mm}\hrulefill

\end{justify}
\end{sizepar}
\vspace*{-4mm}
\caption{Clauses $D1$--$D4$ %(where the ADT variables are underlined) 
are the predicate definitions introduced during transformation. Clauses 11--23
are the clauses derived after transformation from clauses~1--6 and query~7 
of Figure~\ref{fig:ProgCataQuery}. 
\label{fig:TransfProgDefs}}
\end{figure}

%The arguments of predicate \textit{newp} are those of predicate $p$ and those of %the catamorphisms of interest.
Thus, we derive a new version of the initial CHCs 
where each predicate~$p$ has been replaced by the corresponding \textit{newp}.
Then, by applying variants of the fold/unfold transformation rules,
we derive a final, transformed set of CHCs.
When the CHC solver looks for a model of this final set of CHCs, it 
is guided by the fact that suitable constraints, inferred from the query, must hold among
the arguments of the newly introduced predicates, such as \textit{newp}, and thus, the solver can often be more effective.

In our transformation we also introduce, for each predicate \textit{newp}, a predicate called \textit{newp\us woADTs} 
whose definition is obtained by removing the ADT arguments from the definition of \textit{newp}.
For the CHC solvers, it is often easier to find a model for \textit{newp\us woADTs}, rather than for 
$\textit{newp}$, because the solvers
need not handle ADTs at all. However, since each \textit{newp\us woADTs} is an overapproximation of 
$\textit{newp}$, by using the clauses with the ADTs removed, one could wrongly infer unsatisfiability in cases 
when, on the contrary, the initial set of CHCs is satisfiable.

Now, in order to make it easier for the solvers to show satisfiability of sets
of CHCs and, at the same time,
to guarantee the equisatisfiability of the derived set of clauses with respect to % 
the initial set, %among other minor transformations, 
we add to every atom in the body of every derived clause for $\mathit{newp}$ the 
corresponding atom without ADT arguments
%%do the following: (i)~we consider the set $T$ of clauses derived for the $\textit{newp}$ predicates, 
%%and (ii)~we 
%%add to every $\textit{newp}$ atom in the body of a derived clause in $T$ the 
%%to every $\textit{newp}$ atom a \textit{newp\us woADTs} atom}
(see Theorem~\ref{thm:Corr} for the correctness of these atom additions).
By performing these transformation steps starting from clauses~\mbox{1--6} and query~7
(listed in Figure~\ref{fig:ProgCataQuery}) together with the specification
that every variable of 
sort $\mathit{list}(\mathit{int})$ should be associated with a {\it listcount} atom, 
we  derive using our transformation algorithm~\abs (see Section~\ref{sec:Cata})
clauses 11--23 listed in Figure~\ref{fig:TransfProgDefs}. These derived clauses 
%\comment{(together with clauses 5-6 for \textit{listcount})}
are indeed shown to be satisfiable by the Z3 solver, and thus the \textit{Even} property is proved.

 \vspace{-1mm}

\section{CHC Transformation via Catamorphic Abstractions} 
\label{sec:Cata}
In this section we present our transformation algorithm, called~\abs, whose input is: (i)~a set $P$ of
definite clauses, (ii)~a query $Q$ expressing the property to be verified, and (iii)~for each ADT sort, a conjunction of catamorphisms whose 
definitions are included in $P$.
Algorithm \abs~introduces a set of new predicates, 
which incorporate  as extra arguments some 
information coming from the catamorphisms,
%which incorporate information coming from catamorphisms as extra arguments, 
and transforms $P\cup \{Q\}$  into a new set $P'\cup \{Q'\}$ %of clauses
such that $P\cup \{Q\}$ is satisfiable if and only if 
so is $P'\cup \{Q'\}$.

The transformation is effective when the catamorphisms used in the new predicate definitions establish relations that are useful to solve the query. In particular, it is often helpful to use in the new definitions catamorphisms that include the ones occurring in the query, such as the catamorphism $\mathit{listcount}$ 
of our introductory $\mathit{double}$ example.
% of Section~\ref{sec:IntroExample}, 
%the catamorphism used for the transformation is the predicate $\mathit{listcount}$, which is the same catamorphism used in the query.
However, as we will see later, there are cases in which it is important to consider catamorphisms not present in the query (see Example~\ref{ex:quicksort}). The choice of the suitable catamorphisms to be used in the transformation rests upon the programmer's ingenuity {and on her/his understanding of the program behavior. The problem of chosing the most suitable catamorphisms in a fully automatic way is left for future research.}

\subsection{Catamorphic Abstraction  Specifications}\label{subsec:CataAbsSpecif}
The predicates in $P$ different from catamorphisms are called 
\textit{program predicates}. An atom whose predicate is a program predicate is called a
\textit{program atom} and
an atom whose predicate is a catamorphism predicate is called a \textit{catamorphism atom}. 
Without loss of generality, we assume that no clause in $P$ has occurrences of 
both program atoms and catamorphism atoms.
The query $Q$ given in input to \abs~is of the form: 

\smallskip

\hspace{8mm}$\mathit{false} \ \leftarrow c,
\mathit{cata}_1(X,T_1,Y_1), \ldots, \mathit{cata}_n(X,T_n,Y_n), 
\mathit{p}(Z)$
%\comment{MAU: generalizzare a $p_1(...), \ldots, p_k(...)$ ?}

\smallskip

\noindent
where: 
(i)~$\mathit{p}(Z)$ is a program atom and $\mathit{Z}$ is a tuple of 
distinct variables;
(ii)~$\mathit{cata}_1,$ $\ldots,\mathit{cata}_n$ are catamorphism predicates;
{(iii) $c$ is a constraint;}
(iv) $X$ is a tuple of distinct
variables of basic sort;
(v)~${T_1},\ldots,{T_n}$ are ADT variables occurring 
in~$Z$; and
(vi)~{$Y_1,\ldots,Y_n$ are pairwise} disjoint tuples of distinct 
variables of basic sort not occurring in $\mathit{vars}(\{X,Z\})$.
%
%\comment{EMA: definizione incompleta, solo gli autori sanno che $c$ e' un constraint -- commentato nel LaTeX c'\`e  il resto della definizione}
%
%\comment{EMA: uniformare lo stile per rappresentare le tuple: non barrate sopra (questo commento, nella query), barrate sotto (questo commento in cata\_tau) -- update: Y bar \`e locale si pu\`o rimpiazzare con la sequenza Y1,...,Yn}
%;
% and no vi
%(vi)\,$c$~is a constraint such that $\mathit{vars}(c)\cap
%\mathit{vars}(\{X_{1},\!\ldots,\!X_{n},Y_1,\!\ldots,\!Y_n,Z\})\!
%\neq\! \emptyset$.
%
%\comment{era: $\mathit{vars}(c)\!\subseteq\! 
%\mathit{vars}(\{X_{1},$ $\ldots,X_{n},Y_1,\ldots, Y_n, Z\})$. non vale per la query dell'esempio introduttivo}
%$(iv)$~{${Y_1},\ldots,{Y_n}$ are pairwise} disjoint tuples of 
%distinct variables of basic sort
%not occurring in $X_{1},\ldots,X_{n},Z$;
%$(v)$~${T_1},\ldots,{T_n}$ are ADT variables occurring in~$Z$. 
%
Without loss of generality, we assume that the $\mathit{cata}_{i}$'s over 
the same ADT variable are all distinct (this assumption is trivially
satisfied by query~7 of Figure~\ref{fig:ProgCataQuery}).
%(as in query~7 of Figure~\ref{fig:ProgCataQuery}).
%
For each ADT sort $\tau$, a \textit{catamorphic  
abstraction for} $\tau$ is a conjunction of catamorphisms defined as follows:

%%For each sort $\tau$, we define a conjunction of catamorphisms, 
%%referred to as a \comment{\textit{catamorphic %catamorphism-based 
%%abstraction} for the ADT sort $\tau$}, as follows:

\vspace{.5mm}

%\indent
%$\mathit{cata}_\tau(X,T,\overline{Y}) =_\mathit{def} \mathit{cata}_1(X,T,Y_1), \ldots, \mathit{cata}_n(X,T,Y_n)$ 
%\vspace{.5mm}

\indent
$\mathit{cata}_\tau(X,T,Y_{1},\ldots,Y_{n}) =_\mathit{def} \mathit{cata}_1(X,T,Y_1), \ldots, \mathit{cata}_n(X,T,Y_n)$ 
\vspace{.5mm}

\noindent
where: (i)~$T$ is a variable of ADT sort $\tau$, (ii)~$X,Y_{1},\ldots,Y_{n}$ are tuples of variables of basic sort, (iii) the variables in $\{X,Y_{1},\ldots,Y_{n}\}$ are all distinct,
%(iv)~$\overline{Y}$ is a shorthand for $Y_1,\ldots,Y_n$;
and (iv)~the $\mathit{cata_{i}}$ predicates are all distinct.

\indent
Given %a set $P$ of clauses and the 
catamorphic abstractions 
for the ADT sorts~$\tau_{1},\ldots,\tau_{k}$,
%%some suitable catamorphic abstractions 
%%for ADT sorts as we now specify, %$\tau_{1},\ldots,\tau_{k}$, 
a \textit{catamorphic abstraction 
specification for} the set~$P$ of CHCs is a set of expressions, one expression 
%of the form~$(\dagger)$
for each program predicate~$p$ in~$P$ that has at least one argument of ADT sort. The expression for the predicate $p$ is called the \textit{catamorphic abstraction 
specification for~$p$} and it is of the form:

\vspace{.5mm}

\indent
$p(Z) \Longrightarrow \mathit{cata}_{\tau_1}(X,T_1,V_1), 
\ldots, \mathit{cata}_{\tau_k}(X,T_k,V_k)$
%\hfill{$(\dagger)$}\hspace{5mm}

\vspace{.5mm}

\noindent
where: (i)~$Z$ is a tuple of distinct variables,
(ii)~$T_1,\ldots,T_k$ are the distinct variables in~$Z$ 
of {(not necessarily distinct)} ADT
sorts $\tau_1,\ldots,\tau_k$, respectively,
(iii)~$V_1,\ldots,V_k$ are pairwise disjoint tuples of distinct 
variables of basic sort not occurring in $\mathit{vars}(\{X,Z\})$; and\! 
(iv)~$\textit{vars}(X) \cap \textit{vars}(Z)\!=\!\emptyset$. 

%%\comment{Expression~$(\dagger)$ is called the
%%\textit{catamorphic abstraction 
%%specification} for the program predicate~$p$.}

%$p(Z)
%~ \Longrightarrow \mathit{cata}_{\tau_1}(X_1,T_1,\overline{V}_1), \ldots, \mathit{cata}_{\tau_k}(X_k,T_k,\overline{V}_k)$
%
%\vspace{.5mm}
%
%\noindent
%where $T_1,\ldots,T_k$ are the distinct variables in $Z$ of {(not necessarily distinct)} ADT
%sorts $\tau_1,\ldots,\tau_k$, respectively,  
%and $X_1,\ldots,X_k$ are (tuples of)
%variables of basic sort not occurring in $Z$, with $X_i\!=\!X_j$ if and only if $\tau_i\!=\!\tau_j$.  ++++ if?

%+++ the variables $Y_i$ in $\{X_1,\ldots,X_k,\overline{V}_1,\ldots, \overline{V}_k\}$ are
% variables of basic sorts not occurring in $Z$, with $Y_i=Y_j$ if and only if $\tau_i=\tau_j$. 
%
%\comment{*** la variabile X di input deve essere la stessa, ma non quella $\overline{V}$ di output  (o la tupla di quelle di output).}
%%
%%per me:
%%\begin{verbatim}
%%:- spec partition(P,Xs,Ls,Bs) ==> 
%%        listmin(Ls,IsDefMinLs,MinLs), listmax(Ls,IsDefMaxLs,MaxLs),  
%%        listmin(Bs,IsDefMinBs,MinBs), listmax(Bs,IsDefMaxBs,MaxBs), 
%%        listmin(Xs,IsDefMinXs,MinXs), listmax(Xs,IsDefMaxXs,MaxXs),
%%        is_asorted(Xs,B3), is_asorted(Ls,B4), is_asorted(Bs,B5) => constr(true).
%%\end{verbatim}

%\vspace{-1mm}
\begin{example}\label{ex:double} Let us consider our introductory {\it double} example (see Figure~\ref{fig:ProgCataQuery}) and the catamorphic 
%catamorphism-based 
 abstraction for the sort $\mathit{list(int)}$:

\vspace{.5mm}
 $\mathit{cata}_{\mathit{list(int)}}(X,L,N) =_{\mathit{def}} \mathit{listcount}(X,L,N)$ 
 
\vspace{.5mm}
\noindent
This abstraction determines the following catamorphic abstraction specifications for the predicates $\mathit{double}$, $\mathit{eq}$, and
$\mathit{append}$ (and thus, for the set $\{1,\ldots,6\}$ of clauses):

\vspace{.5mm}

\noindent
\hspace{3mm}$\mathit{double}(\mathit{Xs},\mathit{Zs}) %~{\tts{==>}}
 \Longrightarrow \mathit{listcount}(X,\mathit{Xs},N1),\ \mathit{listcount}(X,\mathit{Zs},N2)$

\noindent
\hspace{3mm}$\mathit{eq}(\mathit{Xs},\mathit{Zs}) \Longrightarrow \mathit{listcount}(X,\mathit{Xs},N1),\ \mathit{listcount}(X,\mathit{Zs},N2)$

\noindent
\hspace{3mm}$\mathit{append}(\mathit{Xs},\!\!\mathit{Ys},\!\mathit{Zs}) \Longrightarrow \mathit{listcount}(X,\!\mathit{Xs},\!N1), \mathit{listcount}(X,\!\!\mathit{Ys},\!N2),
\mathit{listcount}(X,\!\mathit{Zs},\!N3)$
%\vspace*{-3mm}

\noindent
{Note that no relationships among the variables $N1$, $N2$ and $N3$ are stated by the specifications. Those relationships will be discovered by the solver after transformation.}~\hfill$\Box$
\end{example}

\vspace*{-3mm}
\begin{example}\label{ex:quicksort}
Let us consider: (i)~a set $\mathit{Quicksort}$ of clauses 
where predicate $\mathit{quicksort}(L,S)$ 
holds if $S$ is obtained from list~$L$ by the quicksort algorithm
% and the elements of $S$ are ordered in weakly ascending order,
%%$S$ is the list ordered in ascending order whose multiset of 
%%elements is the same of that of the list~$L$,
and (ii)~the following query: % \textit{Ord}:
%%\comment{(in this example we 
%%are not interested in stating that $S$ is a permutation of $L$)}:
%%the multisets of elements of 
%%the two lists $L$ and $S$ are equal) 

\vspace{.5mm}
 $\mathit{false} \ \leftarrow\ \mathit{BS}\!=\!\mathit{false},\  
 % \nots \!\mathit{BS},\
\mathit{is\us asorted(S,\!BS)},\  \mathit{quicksort}(L,S)$ \hfill (\!\textit{Ord})\hspace{4mm}

\vspace{.5mm}
\noindent
where $\mathit{is\us asorted(S,\!BS)}$ returns $\mathit{BS}\!=\!\mathit{true}$ 
%(written $\mathit{BS}$) 
if the elements of $S$ are ordered in weakly ascending order, and 
$\mathit{BS}\!=\!\mathit{false}$, %(written $\nots \!\mathit{BS}$)
otherwise. 
The catamorphism 
$\mathit{is\us asorted}$ is defined 
in term of the catamorphism $\mathit{hd}$, as follows:

%we specify below. 
%In the definition of $\mathit{is\us asorted}$, we have that: 
%(i)~if $L$ is not empty, the atom
%$\mathit{hd}(L,\mathit{IsDef},\mathit{Hd})$
%holds if $\mathit{Hd}$ is $\mathit{IsDef}\!=\!\true$ and the head of list~$L$, and (ii)~otherwise, if~$L$ is empty, $\mathit{hd}(L,\mathit{IsDef},\mathit{Hd})$
% holds if $\mathit{IsDef}\!=\!\false$ and $\mathit{Hd}\!=\!0$. If $\mathit{IsDef}\!=\!\false$, then 
%$\mathit{Hd}$ %should 
%\comment{is} not %be 
%used elsewhere in the clause where  $\mathit{hd}(L,\mathit{IsDef},\mathit{Hd})$ occurs.

% for which $\mathit{IsDef}\!=\!\mathit{true}$:  
%whose definition is left to the reader: 

$\mathit{is\us asorted}([\,],\mathit{B}) \leftarrow \mathit{B}\!=\!\mathit{true}$\nopagebreak

$\mathit{is\us asorted}([H | T],\mathit{B})  \leftarrow
   \mathit{B}\!=\!(\mathit{IsDef \Rightarrow (H\! \leq\! HdT\ \&\ BT))}$, \\
\hspace*{20mm}   $\mathit{hd}(T,\mathit{IsDef},\mathit{HdT}),\ 
   \mathit{is\us asorted}(T,\mathit{BT})$ 
   
$\mathit{hd}([\,],\mathit{IsDef},\mathit{Hd}) \leftarrow \ 
\mathit{IsDef} \!=\! \mathit{false}, \ \mathit{Hd}\!=\!0$\nopagebreak
%\nots \!\mathit{IsDef}, \ \mathit{Hd}\!=\!0$

$\mathit{hd}([H|T],\mathit{IsDef},\mathit{Hd}) \leftarrow 
\mathit{IsDef}\!=\!\mathit{true},\ \mathit{Hd\!=\!H}.$

\smallskip
    
%In this example  
\noindent
$\mathit{hd}(L,\mathit{IsDef},\mathit{Hd})$ holds if {\it either} $L$ is the empty list ($\mathit{IsDef}\!=\!\textit{false}$) and $\textit{Hd}$ is $0$ {\it or}
$L$ is a nonempty list ($\mathit{IsDef}\!=\!\textit{true}$)  and $\textit{Hd}$ is its head.
Thus, $\mathit{hd}$ is a total function. Note that the arbitrary value $0$ is not used in the clauses for $\mathit{is\us asorted}$.
%Note that, with reference to Definition~\ref{def:listbtree-cata}, 
%no parameter~$X$ occurs in the list catamorphisms 
%\textit{is\us asorted} and \textit{hd}. 

Let us consider a catamorphic abstraction $\mathit{cata_{\mathit{list(int)}}}$
for the sort $\mathit{list(int)}$, which is the sort of the variables~$L$
and $S$ in $\mathit{quicksort}(L,S)$. That abstraction, consisting of the conjunction of  
 three list catamorphisms $\mathit{listmin}$, $\mathit{listmax}$, and 
$\mathit{is\us asorted}$, is defined as follows:

\vspace{.5mm}
$\mathit{cata_{\mathit{list(int)}}(L,BMinL,MinL,BMaxL,MaxL,BL)}=_{\mathit{def}}$\nopagebreak

\vspace{.5mm}
\hspace*{5mm}
$\mathit{listmin(L,BMinL,MinL),\ listmax(L,BMaxL,MaxL),\ is\us asorted(L,BL)}$

\vspace{.5mm}
\noindent
where: (i)~ if $L$ is not empty, $\mathit{listmin(L,BMinL,MinL)}$ holds if $\mathit{BMinL}\!=\!\mathit{true}$ and $\mathit{MinL}$ is
 the minimum integer in~$L$, and (ii)~otherwise, if $L$ is  empty, 
$\mathit{listmin(L,BMinL,MinL)}$ holds if $\mathit{BMinL}\!=\!\mathit{false}$ and  
$\textit{MinL}\!=\!0$. If $\mathit{BMinL}\!=\!\mathit{false}$, then  $\mathit{MinL}$ should not be used elsewhere in the clause where $\mathit{listmin(L,BMinL,MinL)}$ occurs.
Analogously for $\mathit{listmax}$, instead of  $\mathit{listmin}$.
Then, the
catamorphic abstraction specification for %the predicate 
$\mathit{quicksort}$ is as follows:

\vspace*{.5mm}
\noindent
\hspace*{5mm}$\mathit{quicksort(L,S) \Longrightarrow}$\\
   \hspace*{9mm}$\mathit{listmin(L,\!BMinL,\!MinL),\ listmax(L,\!BMaxL,\!MaxL),\ is\us asorted(L,\!BL)},$\\
   \hspace*{9mm}$\mathit{listmin(S,\!BMinS,\!MinS),\ listmax(S,\!BMaxS,\!MaxS),\ is\us asorted(S,\!BS)}$

%where $\mathit{listmin(L,BMinL,MinL)}$ returns $\mathit{BMinL}\!=\!\mathit{true}$ iff $L$ is not empty, 
%and if $L$ is empty, returns  an arbitrary integer $\textit{MinL}$ if~$L$ is not empty, returns Vthe minimum
%element $\mathit{MinL}$ of $L$, and analogously for $\mathit{listmax}$.  

\vspace*{.5mm}
\noindent
Now, let us assume that in the set of clauses defining $\mathit{quicksort}(L,S)$,
we have the atom $\mathit{partition(\!V\!,\!L,\!A,\!B)}$ that,
%is defined as 
%follows.
% Alb: used V, instead of P, to avoid clash. P was used for the set of clauses.
%occurs in the set $\mathit{Quicksort}$ of clauses. 
given the integer~$V$ and the list~$L$, 
%$\mathit{partition(\!V\!,\!L,\!A,\!B)}$
holds if $\mathit{A}$ is the list made out of the elements of $L$ not larger 
than\,$V$\!, and $B$ is the list made out of the remaining elements of $L$
larger than $V$\!. We have that the 
catamorphic abstraction specification for 
$\mathit{partition}$ which has the list arguments~$L$, $A$, and $B$, is as follows:

\vspace*{.5mm} 
\noindent
\hspace*{5mm}$\mathit{partition(V\!\!,\!L,\!A,\!B) \Longrightarrow}$\\
   \hspace*{9mm}$\mathit{listmin(L,\!BMinL,\!MinL),\ listmax(L,\!BMaxL,\!MaxL),\ is\us asorted(L,\!BL)},$\\
   \hspace*{9mm}$\mathit{listmin(A,\!BMinA,\!MinA),\ listmax(A,\!BMaxA,\!MaxA),\ is\us asorted(A,\!BA),}$\\
   \hspace*{9mm}$\mathit{listmin(B,\!BMinB,\!MinB),\ listmax(B,\!BMaxB,\!MaxB),\ is\us asorted(B,\!BB)}$

\vspace*{.5mm}
\noindent
Note that the catamorphisms \textit{listmin} and \textit{listmax} are not present
in the query \textit{Ord}. % stating that the argument $S$ of \textit{quicksort} is an ordered list. 
However, they are needed for stating the property that, if $\textit{partition}(V\!,\!L,\!A,\!B)$ holds, then
%specifying that for all $V$ and $L$, 
the maximum element of the list $A$ is less than or equal to the minimum element of the list $B$. 
This is a key property useful for proving the orderedness of the list~$S$ %constructed from $L$ 
constructed by $\mathit{quicksort}(L,S)$.
The fact  that the catamorphisms \textit{listmin} and \textit{listmax} are helpful in the proof of the orderedness
of~$S$ rests upon programmer's intuition. However, in our approach the programmer need not explicitly state all the 
properties of \textit{listmin} and \textit{listmax} which are needed for the proof.
Indeed, the relationships among the output variables of \textit{listmin} and \textit{listmax}
are automatically inferred by the CHC solver.~\hfill$\Box$
\end{example}
\vspace*{-4mm}
    
%%
%%We also have the abstraction specification for partition(P,Xs,Ls,Bs) which partitions a list Xs 
%%into the two lists Ls (with elements not larger than P) and Bs (with elements larger than P):
%%partition(P,Xs,Ls,Bs) ==> 
%%        cata_{list(int)}(Xs,BMinXs,MinXs,BMaxXs,MaxXs,BXs),
%%        cata_{list(int)}(Ls,BMinLs,MinLs,BMaxLs,MaxLs,BLs),
%%        cata_{list(int)}(Bs,BMinBs,MinBs,BMaxBs,MaxBs,BBs).
%%\end{verbatim}
%%\end{scriptsize}

% rev(X,Y) ==> length(X,N), elem(K,X,E1), elem(N-K+1,Y,E2) => E1=E2.  
% double-rev(L,RR) ==> elem(K,L,E1), elem(K,RR,E2) => E1=E2.
 %elem should be turned into a total function

\subsection{Transformation Rules} 
\label{subsec:Rules}
%\vspace{-1mm}
The rules for transforming CHCs that use catamorphisms are variants of the usual fold/unfold rules for CHCs~\cite{DeAngelisFGHPP21}.
%, called also 
%Constrained Logic Programs (CLP) in the literature~\cite{DeAngelisFGHPP21}.

\smallskip
\noindent
%\vspace{1mm}
A {\it transformation sequence} from an initial set $S_{0}$ of CHCs to a final set 
$S_{n}$ of CHCs is a sequence 
$S_0 \TransfTo S_1 \TransfTo \ldots \TransfTo S_n$ of sets of CHCs 
such that, for $i\!=\!0,\ldots,n\!-\!1,$ $S_{i+1}$ is derived from~$S_i$, 
denoted
$S_{i} \TransfTo S_{i+1}$, by performing a transformation step consisting in
applying one of the following transformation \mbox{Rules~R1--R5}.

\smallskip
The objective of a transformation sequence constructed by algorithm~\abs~is to derive
from a given set $S_{0}$ a new, equisatisfiable set $S_{n}$ in which for each program predicate~$p$ 
in~$S_{0}$, there is a new predicate \textit{newp} {whose definition is given by 
the conjunction of an atom for $p$ with some catamorphism atoms}. 
With respect to \textit{p}, the predicate \textit{newp} 
has extra arguments that hold the values of the catamorphisms for the arguments of $p$ with ADT sort. 

\smallskip
\noindent
(R1)~{\it Definition Rule.}
Let $D$ be a clause of the form
%$\mathit{newp}(X_1,\ldots,X_k)\leftarrow c,\textit{Catas}, A$, 
$\mathit{newp}(X_1,\ldots,X_k)\leftarrow \textit{Catas}, A$, 
where:
(1)~\textit{newp} is a predicate symbol not occurring in 
the sequence $S_0\TransfTo S_1\TransfTo$ $\ldots\TransfTo S_i$ constructed so far, (2)~$\{X_1,\ldots,X_k\} = \mathit{vars}(\{\textit{Catas},A\})$, 
%(3)~$c$ is a constraint such that 
%$\mathit{vars}(c)\! \subseteq\! \mathit{vars}(\{\textit{Catas},A\})$, 
(3)~\textit{Catas} is a conjunction of catamorphism atoms, with 
$\textit{adt-vars}\/(\textit{Catas})\subseteq \textit{adt-vars}\/(A)$, and
%\comment{\{Adp: in questo lavoro sempre =?\} Mau: no}, and
(4)~$A$ is a program atom.
By the {\it definition introduction rule} we add~$D$ to $S_i$ and we get the new set $S_{i+1}= S_i\cup \{D\}$. 

We will say that $D$ is a definition \textit{for} $A$.
%\vspace{1mm}
%The case when atom $A$ is absent is accommodated by considering % treating it as
%$A$ to be % were 
%$true(X)$, which holds for every $X$ of ADT sort.
% \comment{In that case, for
%Conditions~(2) and (3), we assume that $\mathit{vars}(\{\textit{Catas},A\})= \mathit{vars}(\{\textit{Catas}\})$, and Condition~(4) is vacuously true.}

\smallskip

For any $i\!\geq\!0$, by $\mathit{Defs}_i$ we denote the set of clauses, 
called {\it definitions}, 
introduced by Rule~R1 during the construction of the sequence $S_0\TransfTo 
S_1\TransfTo \ldots\TransfTo S_i$. 
%Thus, $\mathit{Defs}_0\!=\!\emptyset$, and for $j\!=\!0,\ldots, n$, 
%$\mathit{Defs}_j\!\subseteq\!\mathit{Defs}_{j+1}$.
%Note that, by using rules {R2--R6,} one may replace 
%a definition occurring in~$S_j$,
%for some $0\!<\!j\!<\!n$, and hence it may happen 
%that $\mathit{Defs}_{k}\!\not\subseteq\!S_{k}$, for some $k\!>\!j$.

%\smallskip
%\vspace*{-1mm}
\begin{example}
In our \textit{double} example, by applying the definition rule we may introduce the following clause, whose variables of sort $\mathit{list(int)}$ are $B$ and $E$ (the underlining of the list variables~$B$ and $E$
has been omitted here):

\vspace{1mm}
\noindent
$D1$. \ $\mathit{new}1(A,\!B,\!C,\!E,\!F) \leftarrow \mathit{listcount}(A,\!B,\!C),\  
\mathit{listcount}(A,\!E,\!F),\ \mathit{double}(B,\!E)$

\smallskip

\noindent
Thus, $S_1 = S_0 \cup \{D1\}$, where $S_0$ consists of clauses 1--7 of Figure~\ref{fig:ProgCataQuery}.~\hfill$\Box$
\end{example}

%\smallskip

%The clauses for \textit{newp} are obtained by: (i)~first, \textit{unfolding} the definition of \textit{newp}, then 
%(ii)~incorporating some catamorphism atoms and constraints provided by the queries 
%into the clauses derived by unfolding, 
%and (iii)~finally, folding using suitable new definitions.

By making use of the $\mathit{Unf}$ function (see Definition~\ref{def:unf}), we introduce 
the unfolding rule (see Rule R2), which consists of
some unfolding steps followed by the application of the functionality
property, which was presented in previous 
work~\cite{DeAngelisFPP22}. 
Recall that list and binary tree 
catamorphisms and, in general, all catamorphisms are assumed to be 
total functions (see Definition~\ref{def:listbtree-cata}).

\vspace*{-1mm}
\begin{definition}[\mbox{\it{One-step Unfolding}}]\label{def:unf}
Let  $D$$:$ $H\leftarrow c,L,A,R$ be a clause, where~$A$ is an atom,
and let $P$ be a set of definite clauses with 
$\mathit{vars}(D)\cap\mathit{vars}(P)=\emptyset$.
Let $K_{1}\leftarrow c_{1},
B_{1},~\ldots,~K_{m}\leftarrow c_{m}, B_{m}$, with $m\!\geq\!0$,
be the clauses in $P$,
such that, for $j=1,\ldots,m$:
$(\rm{i})$~there exists a most general unifier~$\vartheta_j$ of~$A$ 
and~$K_j$, and {$(\rm{ii})$~the conjunction of constraints $(c, c_{j})\vartheta_j$ is satisfiable.}\\
{\em{One-step unfolding}} produces the following set of CHCs\,$:$

%\smallskip
\vspace{1mm}
$\mathit{Unf}(D,A,P)=\{(H\leftarrow  c, {c}_j,L, B_j, R) 
\vartheta_j \mid  j=1, \ldots, m\}$. 
\end{definition}
\vspace*{-1mm}

\noindent
In the %following Rule R2 and in the 
sequel, 
{\it{Catas}} denotes a conjunction of catamorphism atoms.

\vspace{1mm}
\noindent
(R2)~{\it Unfolding Rule.} 
Let $D$: $\mathit{newp}(U) \If \mathit{Catas}, A$
be a definition in $S_i \cap \textit{Defs}_i$, where $A$ is a program atom, and $P$ be the set of definite clauses in $S_i$.
We derive a new set \textit{UnfD} of clauses by the following three steps.

\smallskip
\noindent
{\it Step}~1.\hspace{1mm}(\!{\it{One-step unfolding of program atom}}) 
$\mathit{UnfD}:=\mathit{Unf}(D,A,P)$;

\smallskip

\noindent
{\it Step}~2.\hspace{1mm}(\!{\it{Unfolding of the catamorphism atoms}})

\hspace{2mm}\begin{minipage}[t]{124mm}
\noindent 
{\bf while} there exists a clause $E$: 
\mbox{$H\! \leftarrow d, {L}, C, {R}$} in $\mathit{UnfD}$, for some conjunctions~$L$ and~$R$ of atoms,
such that $C$ is a catamorphism atom whose argument of ADT sort is not a variable {\bf do}\\
\hspace*{4mm}$\mathit{UnfD}:=(\mathit{UnfD}\setminus\{E\}) \cup \mathit{Unf(E,C,P)}$;~
\end{minipage}

\smallskip

%\rule{0mm}{3mm}
\noindent
{\it Step}~3.\hspace{1mm}(\!{\it{Applying Functionality on catamorphism atoms}})\nopagebreak

\hspace{2mm}\begin{minipage}[t]{114mm}
\noindent
{\bf while} there exists a clause $E$: \mbox{$H \leftarrow d, {L}, \mathit{cata}(X,T,Y1), \mathit{cata}(X,T,Y2),{R}$} 
 in $\mathit{UnfD}$,  for some catamorphism $\mathit{cata}$ {\bf do}\\ 
\hspace*{4mm}$\mathit{UnfD}:=(\mathit{UnfD}-\{E\}) \cup \{H \leftarrow d, Y1\!=\!Y2, {L}, \mathit{cata}(X,T,Y1), {R}\}$.
\end{minipage}

\vspace{2mm}
\noindent
Then, by \textit{unfolding} clause $D$, we get the new set of clauses $S_{i+1}= (S_i \setminus \{D\}) \cup \textit{UnfD}$. 
\begin{example} For instance, in our {\it double} example, by unfolding clause $D1$
we get:

%\smallskip

\vspace*{1mm}
\noindent
$E1$. \ $\mathit{new}1(A,\!B,\!C,\!E,\!F)\! \If\! \mathit{listcount}(A,\!B,\!C),\  
          \mathit{listcount}(A,\!E,\!F),\ \mathit{eq}(B,\!G),$
%          
%\hspace{20mm}
$ \mathit{append}(B,\!G,\!E)$

\vspace*{1mm}

\noindent
Thus, $S_{2}=S_0 \cup \{E1\}$. \hfill$\Box$
\end{example}

%\vspace*{-1mm}
By the following \textit{catamorphism addition rule}, we use the 
catamorphic abstraction specifications for adding catamorphism atoms 
to the bodies of clauses. Here and in what follows, for any two conjunctions $G_1$ and $G_2$ of atoms, 
we say that $G_1$ is a \textit{subconjunction} of $G_2$ if every atom of $G_1$ is an atom of $G_2$.

\smallskip
\noindent
(R3)~{\it{Catamorphism Addition} Rule.}
%Let  $D$: $H\leftarrow c,L,A,R$ be a clause in $S_i$, where~$A$ is a atom,
%and let $P$ be a set of definite clauses with 
%$\mathit{vars}(D)\cap\mathit{vars}(P)=\emptyset$.  be a clause in $S_i = P \cup Q$, where $P$ is a set of definite clauses obtained by 
%applying the unfolding rule, and $Q$ is a
%set of catamorphism-based queries, and 
Let $C$: $H\leftarrow c, \textit{Catas}, A_1,\ldots,A_m$ 
be a clause in $S_i$, 
where $H$ is either \textit{false} or a program atom, and $A_1,\ldots,A_m$ are program atoms.
%whose program atoms are $A_1,\ldots,A_m$. 
Let $E$ be the clause derived from~$C$ as follows:

%\smallskip
%\noindent 
%\textbf{for} each clause $E$: $H\leftarrow e, \textit{Catas}, A_1,\ldots,A_m$ in $P$ {\bf do}

%\newpage
\vspace{0.5mm}
\noindent 
{\bf for} $k=1,\ldots,m$ {\bf do}\nopagebreak

%\noindent
%\hangindent=6mm
%\makebox[2mm]{}\makebox[4mm]{-}let us consider the program atom $A_{k}$; 

\noindent
\hangindent=6mm
\makebox[2mm]{}\makebox[4mm]{-}let $\mathit{Catas}_{k}$ be the conjunction of every catamorphism atom~$F$ 
in $\textit{Catas}$ such that $\mathit{adt\mbox{-}vars}(A_k) \cap 
\mathit{adt\mbox{-}vars}(F) \neq \emptyset$;

\noindent
\hangindent=6mm
\makebox[2mm]{}\makebox[4mm]{-}let $A_k \Longrightarrow \mathit{cata}_1(X,T_1,Y_1),\ldots,
\mathit{cata}_n(X,T_n,Y_n)$ 
be a catamorphic abstraction specification for the predicate of $A_k$,
where the variables in $Y_1,\ldots,Y_n$ do not occur in~$C$, and the conjunction
$\mathit{cata}_1(X,T_1,Y_1),\ldots,$ 
$\mathit{cata}_n(X,T_n,Y_n)$ can be split into two
subconjunctions~$B_1$ and~$B_2$ such that:  

\noindent\hangindent=10mm
\makebox[6mm]{}(i)~a variant $B_1\vartheta$ of $B_1$, for a substitution $\vartheta$ acting 
on $\mathit{vars}(B_{1})$,
is a subconjunction of $\mathit{Catas}_{k}$, and

\noindent\hangindent=10mm
\makebox[6mm]{}(ii)~for every catamorphism atom 
$\mathit{cata}_j(X,T_j,Y_j)$ in $B_2\vartheta$, 
there is no catamorphism atom in $\mathit{Catas}_k$ of the form
$\mathit{cata}_j(V,T_j,W)$ (i.e., there is no catamorphism atom 
with the same predicate acting on the same ADT variable~$T_{j}$);

\noindent
\hangindent=6mm
\makebox[2mm]{}\makebox[4mm]{-}add the conjunction $B_2\vartheta$ to the body of $C$.

\vspace{1mm}

\noindent
Then, by the {\it catamorphism addition rule}, we get the new set $S_{i+1}= (S_i \setminus \{C\}) \cup \{E\}$. 

\vspace*{-1mm}
\begin{example}%For instance, 
\label{ex:R3}
In our {\it double} example, by applying the catamorphism addition rule to clause $E1$, we add the catamorphism $\mathit{listcount}(A,\!H,\!I)$, and we get:

%\smallskip
\vspace{1mm}

\noindent
$E2$. \ $\mathit{new}1(A,\!B,\!C,\!E,\!F) \If \mathit{listcount}(A,\!B,\!C),\ \mathit{listcount}(A,\!E,\!F),\ \mathit{listcount}(A,\!H,\!I),$\nopagebreak

%$E2$. \ $\mathit{new}1(A,\!B,\!C,\!A,\!E,\!F) 
%%\If \mathit{count}(A,\!B,\!C),\ \mathit{count}(A,\!E,\!F),\ 
%%\mathit{count}(A,\!H,\!I),$

\hspace{20mm}
$\mathit{eq}(B,\!H),\ \mathit{append}(B,\!H,\!E)$

\vspace{1mm}
%\smallskip
\noindent
Thus, we get the new set of clauses $S_{3}=S_0 \cup \{E2\}$. \hfill$\Box$
\end{example}

%\vspace*{-1mm}
%%The following \textit{folding rule} allows us to replace a conjunction of
%%a program atom and \comment{one or more DELETE? oppure ***zero or more OPPURE $(n\geq 0)$
%%NOTA: tutti gli atomi vengono foldati} catamorphism atoms by a single atom,
%%whose predicate has been introduced in a previous application of 
%%the Definition Rule. 

The following \textit{folding rule} allows us to replace conjunctions of
catamorphism atoms and program atoms by new program atoms
whose predicates has been introduced in previous applications of 
the definition rule.

\vspace{1mm}
\noindent
(R4)~{\it Folding Rule.}
Let $C$: $H\leftarrow c, \textit{Catas}^{C}, A_1,\ldots,A_m$  
be a clause in $S_i$, where either $H$ is \textit{false} or
$C$ has been obtained by the unfolding rule, possibly followed by the 
application of the 
catamorphism addition rule. $\textit{Catas}^{C}$ is a conjunction of catamorphisms and  $A_1,\ldots,A_m$ are 
program atoms. 
For $k=1,\ldots,m,$

\smallskip
\noindent
\begin{minipage}{122mm}
\noindent
\hangindent=3mm
\makebox[3mm]{-}let $\mathit{Catas}_{k}^{C}$ be the conjunction of every catamorphism atom $F$ 
in $\textit{Catas}^{C}$ such that $\mathit{adt\mbox{-}vars}(A_k) \cap 
\mathit{adt\mbox{-}vars}(F) \neq \emptyset$;
\end{minipage}

\noindent
\begin{minipage}{122mm}
\noindent\hangindent=3mm
\makebox[3mm]{-}let
$D_k$: $H_k \leftarrow \mathit{Catas}^{D}_{k}, A_k$ be a clause in $\mathit{Defs}_i$ (modulo variable renaming)
such that $\mathit{Catas}^{C}_{k}$ is a %\bcomment{(possibly non-proper)} 
subconjunction of $\mathit{Catas}^D_{k}$.
\end{minipage}

\smallskip\noindent
Then, by \textit{folding \( C\)
	using~\( D_1,\ldots,D_m\)}, we derive clause 
\(E  \):~\( H\leftarrow c, H_1,\ldots,H_m \), and we get the new set of clauses
\( S_{i+1}= (S_{i}\setminus\{C\})\cup \{E \} \).

%\vspace{-1mm}
\begin{example}\label{ex:E2-fold-E3} 
In order to fold clause $E2$ (see Example~\ref{ex:R3}) according to the folding rule~R4, we introduce 
for the program atoms $\mathit{append}(B,H,E)$ 
and $\mathit{eq}(B,H)$ that occur in the body of $E2$, the new definitions~$D2$ and $D3$, respectively.
Those new definitions are shown in %the upper part of 
Figure~\ref{fig:TransfProgDefs}.
%%occurring in the body of clause~$E2$, we introduce the claus
%%clause~$D3$ is for the program atom $\mathit{eq}(B,H)$, both atoms occurring 
%%in the body of clause~$E2$. 
Then, by folding clause~$E2$ using $D2$ and $D3$, we get: 

% -fold-   
%new1(A,B,C,D,E,F) :- A=D, A=G, C=H, D=A, D=I, I=A, I=J, K=L, 
%          new2(I,M,K,D,E,F,A,B,C), new3(J,M,L,G,B,H).
\vspace*{.5mm}
%\smallskip
\noindent
$E3$. \ $\mathit{new}1(A,\!B,\!C,\!E,\!F) \leftarrow
 \mathit{new}2(A,\!M,\!K,\!E,\!F,\!B,\!C),\ \mathit{new}3(A,\!M,\!K,\!B,\!C)$.

\vspace*{.5mm}
\noindent
Also, query $7$ (see Figure~\ref{fig:ProgCataQuery}) can be folded using definition~$D1$, and we get:

\vspace*{1mm}
\noindent
$E4$. \ $\mathit{false} \leftarrow C\!=\!2\,D\!+\!1,\ 
\mathit{new}1(A,\!E,\!F,\!G,\!C)$ 

\vspace*{1mm}

\noindent
Thus, $S_{4}=(S_0 \setminus \{7\}) \cup \{E3,E4,D2,D3\}$.
Then, we will continue by transforming the newly introduced definitions~$D2$ and~$D3$.~\hfill$\Box$
\end{example}

%\vspace*{-2mm}

The following Rule R5 is a new transformation rule that allows us: (i)~to introduce 
new predicates by erasing ADT arguments from existing predicates, 
and (ii)~to add atoms with these new predicates to the body of a clause.

\smallskip
\noindent
(R5)~{\it Erasure Addition Rule.} 
% Alberto:
% in rule R5 we could say that A is a "program atom" because in the transformation system also
% the catamorphisms (come "count") diventano nel final program dei newk atomi, cioe' dei program atoms. 
% But we cannot do so because the "simpler" presentation of Figure 2 where the newk which are 
% the same as catamorphisms regain the name of the catamorphisms (see count).
Let $A$ be the atom $p(t_1,\ldots,t_k,u_1,\ldots,u_m)$, where $t_1,\ldots,t_k$ have 
(possibly distinct) basic sorts and $u_1,\ldots,u_m$ have (possibly distinct) ADT 
sorts. 
We define the \textit{ADT-erasure} of 
$A$, denoted $\chi_{\!\mathit{wo}}(A)$, to be the atom $p\us{\mathit{woADTs}}(t_1,\ldots,t_k)$, where $p\us{\mathit{woADTs}}$ is a 
new predicate symbol.
Let $C$: $H\If c, A_1,\ldots,A_n$ be a clause in $S_i$. \\
%where $H$ is not \textit{false},  
%\comment{variables or terms? in the proof are variables.}
Then, by the \textit{erasure addition rule}, from $C$ we derive the two new clauses:

%\smallskip
\vspace{1mm}

%\noindent

\makebox[60mm][l]{$\chi_{\!\mathit{wo}}(H)\!\If\! c, \chi_{\!\mathit{wo}}(\!A_1\!),\ldots,\chi_{\!\mathit{wo}}(\!A_n\!)$,} \makebox[30mm][l]{denoted $\chi_{\!\mathit{wo}}(C)$,} and 

\vspace{1mm}
\makebox[60mm][l]{$H\!\If\! c, A_1,\chi_{\!\mathit{wo}}(\!A_1\!),\ldots,A_n, \chi_{\!\mathit{wo}}(\!A_n\!)$,} denoted $\chi_{\!\mathit{w\&wo}}(C)$,

%\smallskip
\vspace{1mm}
\noindent
and we get the new set of clauses

$S_{i+1}=  \{\chi_{\!\mathit{w\&wo}}(C) \mid C\in S_i\}\ \cup\ \{\chi_{\!\mathit{wo}}(C) \mid C$ is a clause in $S_i$ whose head is not \textit{false}\}.

%\vspace*{-1mm}
\begin{example} Let us consider clause $E3$ of Example~\ref{ex:E2-fold-E3}.
We have that: 

\makebox[49mm][l]{$\chi_{\!\mathit{wo}}(\mathit{new}1(A,B,C,E,F))$} $=$~ $\newiwoADTs(A,C,F)$,

\makebox[49mm][l]{$\chi_{\!\mathit{wo}}(\mathit{new}2(A,M,K,E,F,B,C))$} $=$~ $\newiiwoADTs(A,K,F,C)$, 
 
\makebox[49mm][l]{$\chi_{\!\mathit{wo}}(\mathit{new}3(A,M,K,B,C))$} $=$~ $\newiiiwoADTs(A,K,C)$.

%\indent
%\comment{\makebox[49mm][l]{$\chi_{\!\mathit{wo}}(\mathit{listcount}(A,L,B))$} $=$~ $\listcountwoADTs(A,B)$.}
  
\noindent      
Thus, from clause $E3$, by erasure addition we get clauses~12 and 18 of 
Figure~\ref{fig:TransfProgDefs}.~\hfill$\Box$
\end{example}

The following theorem is a consequence of well-known results for CHC
transformations (see, for instance, the papers cited in a recent survey~\cite{DeAngelisFGHPP21}). 

%%Note that, in particular, the correctness of Rule R3 easily follows
%%from the fact that catamorphisms are total functions.
%%The correctness of Rule~R5 follows from the fact that, for $i\!=\!
%%1,\ldots,n$,
%%atom $\chi(A_{i})$ is an overapproximation of $A_{i}$ \comment{(thus, 
%%$A_{i}$ implies $\chi(A_{i})$)},
%%and hence the addition of \comment{$\chi(A_{i})$ to the body of 
%%a clause that includes $A_{i}$} does not affect 
%%\comment{clause satisfiability}.

%%and hence the addition of atom $\chi(p)(X)$ to the body of 
%%a clause \comment{that includes $p(X)$} does not affect \comment{satisfiability}.

%\vspace{-2mm}
\begin{theorem}[Correctness of the Rules]
\label{thm:Corr}
Let $S_0 \TransfTo S_1 \TransfTo \ldots \TransfTo S_n$ be 
a transformation sequence
using Rules {\rm{R1--R5}}.
Then, $S_0$ is satisfiable if and only if $S_n$ is satisfiable.
\end{theorem}

\begin{proof} 

The proof consists in showing that Rules R1--R5 presented earlier in this section can be derived from the transformation rules considered in previous work~\cite{De&22a} and proved correct based on results by Tamaki and Sato~\cite{TaS86} for logic programs and Etalle and Gabbrielli~\cite{EtG96} for constraint logic programs. 
Below we will recall these transformation rules.

Let us first introduce the notion of {\it stratification} for a set of clauses~\cite{Llo87}. 
Let~$\mathbb N$ be the set of the natural numbers and {\it Pred} be the set of the predicate names.
A \emph{level mapping} is a 
function~$\lambda\!:\mathit{Pred}\!\rightarrow\!\mathbb{N}$. For every predicate $p$,
the natural number $\lambda(p)$  is said to be the {\it level\/} of~$p$.
Level mappings are extended to atoms by stating that 
the level $\lambda(A)$ of an atom $A$ is the 
level of its predicate symbol.
A clause \( H\leftarrow c, A_{1}, \ldots, A_{n}
\) is {\it stratified with respect to}~$\lambda$ if either $H$ is \textit{false} or, 
for \( i\!=\!1,\ldots ,n \),
\(\lambda (H)\geq \lambda(A_i)\).
A set $P$ of CHCs is {\it stratified~with respect to $\lambda$} if all clauses
of $P$ are stratified with respect to~$\lambda$.
%Clearly, for every set~$P$ of CHCs, there exists a level mapping~$\lambda$ such that
%$P$ is stratified with respect to $\lambda$~\cite{Llo87}. 

%\comment{REPLACE $\Rightarrow$ by $\TransfTo$  }
\medskip

A {\it DUFR-transformation sequence from} $S_{0}$ {\it to} $S_{n}$
is a sequence 
\mbox{$S_0 \TransfTo S_1 \TransfTo \ldots \TransfTo S_n$} of sets of CHCs 
such that, for $i\!=\!0,\ldots,n\!-\!1,$ $S_{i+1}$ is derived from $S_i$, denoted
\mbox{$S_{i} \TransfTo S_{i+1}$}, by
applying one of the following rules: (i)~Rule~D, (ii)~Rule~U, (iii)~Rule~F, and (iv)~Rule~G. 
(To avoid confusion with Rules R1--R5 presented earlier in this section, in this proof we use 
the letters D, U, F, and G  to identify the rules presented in previous work \cite{De&22a}.)
We assume that the initial set
$S_0$ is stratified with respect to~a given level mapping~$\lambda$.

\bigskip
%The Definition Rule allows us to introduce new predicate definitions.
%\medskip
%\hrule
%\vspace*{1.5mm}
%\smallskip
\noindent
{\bf (Rule D)}  
Let $D$ be the clause $\textit{newp}(X_1,\ldots,X_k)\leftarrow 
c,A_1,\ldots,A_m$, where:
(1)~\textit{newp} is a predicate symbol not occurring in 
the sequence $S_0\TransfTo S_1\TransfTo\ldots\TransfTo S_i$ constructed 
so far, \mbox{(2)~$c$ is a constraint,} 
(3)~the predicate symbols of $A_1,\ldots,A_m$
occur in $S_0$, and 
(4)~$\{X_1,\ldots,X_k\}\subseteq \mathit{vars}(\{c,A_1,\ldots,A_m\})$.
Then, by Rule D, % by {\it definition}
 we get $S_{i+1}= S_i\cup \{D\}$. 
We define the level mapping $\lambda$ of \textit{newp} to be equal 
to $\textit{max}\,\{\lambda(A_i) \mid i=1,\ldots,m\}$.

%\smallskip
%\hrule

\medskip

For any $i\geq 0$, we denote by $\textit{Defs}_i$ the set of clauses 
introduced by Rule~D during the construction of
$S_0\TransfTo S_1\TransfTo\ldots\TransfTo S_i$. 
%Thus, $\textit{Defs}_0\!=\!\emptyset$, and for $j\!=\!0,\ldots, n,$ 
%$\textit{Defs}_j\!\subseteq\!\textit{Defs}_{j+1}$.
%Note that, by using rules (U), (F), (Fun), (Tot), one may replace 
%a definition occurring in~$S_h$,
%for some $0\!<\!h\!<\!n$, and hence it may happen 
%that $\textit{Defs}_{k}\!\not\subseteq\!S_{k}$, 
%for some $k$ such that $h\!<\!k\!\leq\!n$.
%for some $k\!>\!h$.

\bigskip
Rule U consists in an application of the one-step unfolding of Definition~\ref{def:unf}.

\medskip

%\hrule
%\vspace*{1.5mm}
%\smallskip
\noindent
{\bf (Rule U)} %~Unfolding Rule.} 
Let  $C$: $H\leftarrow c,G_L,A,G_R$ be a clause in $S_i$, where $A$ is an atom.
%Without loss of generality, we assume that
%$\mathit{vars}(C)\cap\mathit{vars}(S_0)=\emptyset$.
%Let {\it Cls}: $\{K_{1}\leftarrow c_{1},
%B_{1},~\ldots,~K_{m}\leftarrow c_{m}, B_{m}\}$, with $m\!\geq\!0$,
%be the set of clauses in $S_0$,
%such that: for $j=1,\ldots,m$,
%(1)~there exists a most general unifier~$\vartheta_j$ of $A$ and
%$K_j$, and {(2)~the conjunction of constraints $(c, c_{j})\vartheta_j$ is satisfiable.}
%Let $\mathit{Unf}(C,A,S_0)$ be the set $\{(H\leftarrow  c, {c}_j,G_L, B_j, G_R) 
%\vartheta_j \mid  j=1, \ldots, m\}$ of clauses.
%
Then, by applying Rule U to $C$ 
%{\it unfolding~$C$ 
with respect to $A$, %we derive the set  $\mathit{Unf}(C,A,S_0)$
we get $S_{i+1}= (S_i\setminus\{C\}) \cup \mathit{Unf}(C,A,S_0)$.
%\smallskip
%\hrule

\bigskip
\noindent
{\bf (Rule F)} %~Folding Rule.} 
Let $C$: $H\leftarrow c, G_L,Q,G_R$ be a clause in $S_i$, 
and let
$D$: $K \leftarrow d, B$ be a variant of a clause in $\textit{Defs}_i$.
%such that $\mathit{vars}(C)\cap\mathit{vars}(D)=\emptyset$.
Suppose that:
(1)~either $H$ is $\mathit{false}$ or \mbox{$\lambda(H) \geq \lambda(K)$,} and
(2)~there exists a substitution~$\vartheta$ such that~\mbox{$Q\!=\! B\vartheta $} and
$\mathbb D\models \forall(c \rightarrow d\vartheta)$.
Then, by applying Rule F to $C$ using $D$, %\textit{folding \( C\)\,using definition\,\( D\)}, 
we derive clause 
\(E  \):~\( H\leftarrow c, G_{\!L}, K\vartheta, G_{\!R} \), and we get
\( S_{i+1}= (S_{i}\setminus\{C\})\cup \{E \} \).

\bigskip

In the next Rule R, called \emph{goal replacement}, and in the rest of the proof, by $\textit{Definite}(S_0)$ we denote the set of definite clauses belonging to $S_0$.

\medskip

\noindent 
\textbf{(Rule R)} %~Goal Replacement Rule.} 
Let  
$C$:~${H}\leftarrow c, c_{1}, {G}_{L},
{G}_{1}, {G}_{R}$ be a clause in~${S}_{i}$.
Suppose that the following two conditions hold:

(R.1)~$M(\textit{Definite}(S_0)\cup \mathit{Defs}_i) \models \forall\,( (\exists T_1.\, c_{1}\! \wedge \! {G}_{1})
\leftrightarrow (\exists T_2.\, c_{2}\! \wedge \! {G}_{2}))$, and

(R.2)~either $H$ is $\mathit{false}$ or, for every atom $A$ occurring 
in $G_2$ and not in $G_1$, 
$\lambda(H)\!>\!\lambda(\mathit{A})$

\noindent
where: 

%
%\vspace*{1mm}
$T_1 = \mathit{vars}(\{c_{1},{G}_{1}\}) 
\setminus \mathit{vars}(\{{H}, c, {G}_{L},{G}_{R}\})$,~~ and

$T_2 = \mathit{vars}(\{c_{2},{G}_{2}\}) 
\setminus \mathit{vars}(\{{H}, c, {G}_{L},{G}_{R}\})$. 
%of variables
%which will denoted by the tuple $T$ of variables in the following
%Conditions (W.1) and (S).  

\noindent
Then, by Rule R, in 
clause~$C$ we  \emph{replace} $c_1,G_1$ \emph{by} $c_2,G_2$, and we derive clause~$D$:
${H}\leftarrow c, c_{2}, {G}_{L},
{G}_{2}, {G}_{R}$. We get 
$S_{i+1} = (S_{i}\setminus\{C\})\cup \{D\}$. 

\medskip

The following result guarantees that, for any DUFR-transformation sequence $S_0 \TransfTo S_1 \TransfTo \ldots \TransfTo S_n$ satisfying Condition~(C),
$S_0$ and $S_n$ are equisatisfiable~\cite{TaS86,EtG96,De&22a}.

%\comment{First??, we have the following result.} 

%%The soundness of the transformation rules is guaranteed by the 
%%following result whose proof is given in the Appendix. 

\begin{theorem}[Correctness of the DUFR-Transformation Rules]
\label{thm:CorrOfUF}
Let $S_0 \TransfTo S_1 \TransfTo \ldots \TransfTo S_n$ be 
a DUFR-transformation sequence.
Suppose that the following condition holds:
	
\vspace{.5mm}
\noindent\hangindent=8mm
\makebox[8mm][l]{\rm \,(C)}for $i \!=\! 1,\ldots,n\!-\!1$, if $S_i \TransfTo S_{i+1}$ by 
folding a clause in $S_i$ using a definition $D\!: H \leftarrow c,B$  
in $\mathit{Defs}_i$, 
then, for some $j\! \in\!\{1,\ldots,i\!-\!1,i\!+\!1,\ldots, n\!-\!1\}$, 
$ S_{j}\TransfTo S_{j+1}$ by unfolding~$D$ 
with respect to an atom $A$ such that $\lambda(H)=\lambda(A)$. 
	
\vspace{.5mm}
\noindent
Then, 

(1) for $i=1,\ldots,n,$ $M(\mathit{Definite}(S_0)\cup \mathit{Defs}_i) = M(\mathit{Definite}(S_i)),$  and

(2) $S_0$ is satisfiable if and only if $S_n$ is satisfiable.
\end{theorem}

\noindent
Now, we will show that each application of Rules R1--R5 can be obtained by one or more applications of Rules D, U, F, R. Furthermore, for any transformation sequence
$S_0 \TransfTo S_1 \TransfTo \ldots \TransfTo S_n$ constructed using Rules R1--R5, there exists a DUFR-transformation sequence $S_0 \TransfTo T_0 \TransfTo \ldots \TransfTo T_r  \TransfTo S_n$ satisfying Condition (C) of Theorem \ref{thm:CorrOfUF}.

In order to recast Rules R1--R5 in terms of Rules D, U, F, and R, we first %need to 
introduce a suitable level 
mapping $\lambda$ defined as follows: for any predicate $q$, (i)~$\lambda(q)\!= \!2$, if $q$ is a program predicate of the initial set of clauses or a new program predicate introduced by Rule R1, 
and (ii)~\mbox{$\lambda(q)\!= \!1$}, if $q$ is a catamorphism predicate, and (iii)~$\lambda(q)\!= \!0$, if $q$ is a 
new predicate 
symbol introduced by Rule R5. We have that the initial set $S_0$ of CHCs is stratified with respect to~$\lambda$.
Let us first consider the four Rules R1--R4.

\begin{itemize}

\item Rule R1 is a particular case of Rule D, where in the body of clause $D$, (i)~the constraint~$c$ is absent, (ii)~exactly one atom among $A_1,\ldots,A_m$ is a program atom, (iii)~all other atoms are catamorphism atoms, and (iv)~$\{X_1,\ldots,X_k\}= \mathit{vars}(\{A_1,\ldots,A_m\})$. By our definition of the level mapping, $\lambda(\mathit{newp})=2$, as one of the $A_i$'s is a program atom.

\item Rule R2 consists of applications of Rules U and R. Indeed, in R2, (i) Steps 1 and 2 are applications of Rule U where $P$ is $S_0$, and (ii) Step 3 is an application of Rule R. To see Point (ii), note that every catamorphism $\mathit{cata}$ is, by definition, a functional predicate (see Section~\ref{sec:CHCs}), and hence
$M(\mathit{Definite}(S_0))\models \forall (\mathit{cata}(X,T,Y1) \wedge \mathit{cata}(X,T,Y2) \rightarrow Y1\!=\!Y2) $. Thus, for any $i\geq 0$, 

\smallskip

$M(\mathit{Definite}(S_0) \cup \mathit{Defs}_i)\models \forall (\mathit{cata}(X,T,Y1) \wedge \mathit{cata}(X,T,Y2) \leftrightarrow Y1\!=\!Y2 \wedge \mathit{cata}(X,T,Y1)) $

\smallskip

that is, Condition (R.1) of Rule R holds. Also Condition (R.2) holds, as the head $H$ of the clause has a predicate $\mathit{newp}$ introduced by definition, and hence $\lambda(\mathit{newp})=2$, while we have stipulated that $\lambda(\mathit{cata})=1$.

\item Rule R3 consists of applications of Rule R. Indeed, R3 adds to the body of a clause~$C$ (zero or more) catamorphism atoms $\mathit{cata}_j(X,T_j,Y_j)$ such that no variable in the tuple~$Y_j$ occurs in $C$. 
The assumption that catamorphisms are total functions enforces that 
$M(\mathit{Definite}(S_0)) \models \forall X, T_j\ \exists Y_j.\ \mathit{cata}_j(X,T_j,Y_j)$, and hence

\smallskip

$M(\mathit{Definite}(S_0) \cup \mathit{Defs}_i)\models \forall ( true \leftrightarrow \exists Y_j.\ \mathit{cata}_j(X,T_j,Y_j))$

\smallskip

that is, Condition (R.1) of Rule R holds. Also Condition (R.2) holds, as the head $H$ of clause $C$ is either $\mathit{false}$ or a program atom. In the latter case $\lambda(H)=2$, while we have stipulated that $\lambda(\mathit{cata}_j)=1$.

\item
Rule R4 consists of applications of Rules R and F. 
%%%Indeed, an application of Rule R3 can be 
%%%as the result of the application of the following for-loop of applications of Rules R and F:
%%%%
%%% Indeed, the result of applying Rule R3 can be 
%%%obtain by the following for-loop of applications of Rules R and F:
%%%%
Indeed, an application of Rule R3 is equivalent to the following for-loop of applications of Rules R and F:
 for \mbox{$k\!=\!1,\ldots, m$}, first, 
(i)~the addition of the catamorphism atoms occurring (modulo variable renaming) in $\mathit{Catas}^{D}_{k}$ and not in its subconjunction $\mathit{Catas}^{C}_{k}$ (as mentioned above, this catamorphism addition is an instance of Rule R), and then, (ii)~the application of Rule F, thereby replacing the conjunction  
$(\mathit{Catas}^{D}_{k}, A_k)$ by $H_k$. 
 
\end{itemize}

Therefore, for any transformation sequence $S_0 \TransfTo S_1 \TransfTo \ldots \TransfTo S_i$ constructed using 
Rules~\mbox{R1--R4}, there exists a DUFR-transformation sequence $S_0 \TransfTo T_0 \TransfTo \ldots \TransfTo T_r  \TransfTo S_i$. When applying Rule R4 to a clause $C$ during the construction of $S_0 \TransfTo S_1 \TransfTo \ldots \TransfTo S_i$, either the head of $C$ is \textit{false} or
$C$ has been obtained by the unfolding rule (possibly followed by 
catamorphism addition). This implies that in $S_0 \TransfTo T_0 \TransfTo \ldots \TransfTo T_r  \TransfTo S_i$ we have that Condition~(C) of Theorem~\ref{thm:CorrOfUF} holds. Thus,
by Theorem~\ref{thm:CorrOfUF} we get: $M(\mathit{Definite}(S_0)\cup \mathit{Defs}_i) = M(\mathit{Definite}(S_i))$. 

Now, suppose that we apply Rule R5 to the set $S_i$ of clauses. We have that, for every  predicate $p$ occurring in $S_i,$

\vspace{1mm}
$M(\!\mathit{Definite}(S_i) \cup \chi_{\!\mathit{wo}}(S_i))\models
\forall (p(X_1,\sdots,\!X_k,\!Y_{\!1},\sdots,\!Y_m) \rightarrow p\us{\mathit{woADTs}}(X_1,\sdots,\!X_k))$~~~\hfill$(\dagger)$

\vspace{1mm}
\noindent
where $\chi_{\!\mathit{wo}}(S_i) = \{\chi_{\!\mathit{wo}}(C) \mid C$ is a clause in $S_i$ whose head is not \textit{false}$\}$. 
Now, it is the case that an application of Rule R5 is realized by a sequence of applications of Rule R. Indeed, for each addition of an atom $p\us{\mathit{woADTs}}(t_1,\ldots,t_k)$ to the body of a clause~$C$ by~R5,
Condition (R.1) holds, as the above relation~$(\dagger)$ is equivalent to:

\vspace{1mm}
$M(\mathit{Definite}(S_i) \cup \chi_{\!\mathit{wo}}(S_i))\models$\nopagebreak

~$\forall (p(X_1,\ldots\!,X_k,Y_1,\ldots\!,\!Y_m) \leftrightarrow (p(X_1,\ldots\!,X_k,Y_1,\ldots\!,\!Y_m)  \wedge p\us{\mathit{woADTs}}(X_1,\ldots,X_k)))$
 
\vspace{1mm}
\noindent
and $M(\mathit{Definite}(S_i)\cup \chi_{\!\mathit{wo}}(S_i)) = M(\mathit{Definite}(S_0\cup \chi_{\!\mathit{wo}}(S_i)) \cup \mathit{Defs}_i)$, because the predicates in $\chi_{\!\mathit{wo}}(S_i)$ do not occur in $S_0,\ldots,S_i$.
Also Condition (R.2) holds, because the head $H$ of $C$ is either \textit{false} or $\lambda(H)\!\geq\! 1$ and $\lambda(p\us{\mathit{woADTs}})\!=\!0$.

%\smallskip
Therefore, %we conclude that,  %In summary, we have shown that, 
for any transformation sequence $S_0 \TransfTo S_1 \TransfTo \ldots \TransfTo S_n$ constructed using Rules R1--R5, there exists a DUFR-transformation sequence $S_0 \TransfTo T_0 \TransfTo \ldots \TransfTo T_r  \TransfTo S_n$.
Then, by Theorem~\ref{thm:CorrOfUF}, we get that $S_0$ is satisfiable if and only if $S_n$ is satisfiable. \hfill\end{proof}

\subsection{The Transformation Algorithm \abs} 
\label{subsec:Strategy}
The set of the new predicate definitions needed during the execution of the transformation 
algorithm~\abs\ is not given in advance. In general, that set
depends on: (i)~the initial set~$P$ of CHC clauses, (ii)~the given query $Q$
specifying the property of interest to be proved, and (iii)~the given  
catamorphic abstraction specification $\alpha$ for $P$.
As we will see, we may compute that set of new definitions as the least fixpoint of an operator,
called $\tau_{P\cup\{Q\},\alpha}$, which transforms a given set~$\Delta$ of
predicate definitions into a new set $\Delta'$ of predicate definitions. 
First, we need the following notions.

Two definitions $D_1$ and $D_2$ are said to be \emph{equivalent}, denoted $D_1\equiv D_2,$ if they can be made
identical by performing the following transformations: (i)~renaming of the head predicate, 
(ii)~renaming of the variables, (iii)~reordering of the variables in the head, and (iv)~reordering of the atoms in the body. 
We leave it to the reader to check that the results presented in this section are indeed independent of the choice 
of a %one 
specific definition in its equivalence class. %among all equivalent ones.

A set $\Delta$ of definitions is said to be 
\emph{monovariant} if, for each program predicate $p$, in $\Delta$
there is at most one definition  having an occurrence of $p$ in its body.
The transformation algorithm \abs~and the operator $\tau_{P\cup\{Q\},\alpha}$ work on monovariant sets of definitions and are defined by means of the $\mathit{Define}$, $\mathit{Unfold}$, $\mathit{AddCata}$, $\mathit{Fold}$, and $\mathit{AddErasure}$ functions defined in Figure~\ref{fig:Functions}.

% -----------------------------------------------------------------------------
% -----------------------------------------------------------------------------
\begin{figure}[!ht]
%\begin{figure}
\begin{sizepar}{9.3}{11}
\begin{minipage}{125mm}
\noindent \hrulefill 

\noindent 
Given a set $\mathit{Cls}$ of clauses and a monovariant set $\Delta$ of definitions,\\
$\mathit{Define}(\mathit{Cls},\Delta)$ returns a monovariant set $\Delta'$ of new definitions computed as follows.

\vspace{1mm}

\noindent $\Delta' := \Delta$;

\noindent 
{\bf for} each clause $H\leftarrow c, G$ in $\mathit{Cls}$, 
where goal $G$ contains at least one ADT variable {\bf do}

\noindent 
\hspace{3mm}{\bf for} each program atom $A$ in $G$ {\bf do}
%\comment{rivedere indentazione}

\smallskip
\hspace{6mm}
\begin{minipage}{116mm}
\noindent
%let $\mathit{Catas}\!_A$ be the conjunction of every catamorphism atom $F$ in $G$ such that $\mathit{adt\mbox{-}vars}(F) \subseteq 
%\mathit{adt\mbox{-}vars}(A)$

%$\mathit{Catas}\!_A:=\{ F \,|\, F $ is a catamorphism atom in $G$ and 
%$\mathit{adt\mbox{-}vars}(F)\subseteq\mathit{adt\mbox{-}vars}(A) \}$;

$\mathit{Catas}\!_A:=\{ F \,|\, F$ is a catamorphism atom in $G$ and $\mathit{adt\mbox{-}vars}(F)\subseteq\mathit{adt\mbox{-}vars}(A) \}$;

\smallskip
{\bf if} there is a clause %\comment{\{equivalent to ? come gestire l'equivalenza? \}}
$\mathit{D}\!\!:$ $\mathit{newp}(U) \If B,A$ in $\Delta'$, for some conjunction $B$ of catamorphism atoms %{\bf then}

\smallskip
\begin{minipage}{115mm}
{\bf then if} $\mathit{Catas}\!_A$ is {\em not} a 
subconjunction of $B$ %ALB: (and $\mathit{Catas}\!_A\!\not =\! B$) 
{\bf then} \hfill // \makebox[12mm][l]{({\bf Extend})}

\hspace{9mm}\hangindent=9mm
{\rm by applying the definition rule R1, introduce the definition}\\ %\underline{\Down introduce definition} 
$\mathit{ExtD}$: $\mathit{extp}(V) \If B', A$,
where: 
(i) $\mathit{extp}$ is a new predicate symbol,
%(ii) $V\!=\!\mathit{vars}(\{B',A\})$, and 
(ii)~$B'$ is the conjunction of the distinct catamorphism atoms occurring either in~$B$ or in $\mathit{Catas}\!_A$, and 
(iii) $V\!=\!\mathit{vars}(\{B',A\})$;

\hspace{9mm}$\Delta' := (\Delta' \setminus \{D\}) \cup \{\mathit{ExtD}\}$;

\end{minipage}

\smallskip
\begin{minipage}{113mm}
%%{\bf else} \hfill //  \makebox[15mm][l]{({\bf Add})}

\hangindent=7mm
{\bf else} {\rm by applying the definition rule R1, introduce the definition} \hfill // \makebox[12mm][l]{({\bf Add})}\\ %\underline{\Down introduce definition} 
$\mathit{NewD}$: $\mathit{newp}(V) \If  \mathit{Catas}\!_A, A$,
where: 
(i)~$\mathit{newp}$ is a new predicate symbol, and
(ii)~$V\!=\!\mathit{vars}(\{\mathit{Catas}\!_A, A\})$;

\hspace{7mm}$\Delta' := \Delta' \cup \{\mathit{NewD}\}$;

\smallskip
\end{minipage}

\end{minipage}%%%%%%%%%%%%%%%%%%%%%%%

\noindent \hrulefill

\noindent 
Given a set $\Delta\!=\!\{D_i \mid 0\!\leq\!i\!\leq\!n\}$ of definitions
%Given a set $\Delta\!=\!\{D_1,\ldots,D_n\}$, with $n\!\geq\!0$, of definitions 
and a set $P$ of definite clauses,\\
$\mathit{Unfold}(\Delta,P)\!=\! \bigcup^n_{i=1} \mathit{UnfD}_i$, where $\mathit{UnfD}_i$ is the set of clauses derived by 
%applying the\linebreak \hspace*{41mm}unfolding rule R2 to clause~$D_i$.
applying the unfolding rule R2 to clause~$D_i$.

\vspace*{-2mm} \noindent \hrulefill

%\comment{
\noindent
%Given a set $\mathit{Cls}\!=\!\{C_1,\!\ldots,\!C_n\}\!$, with $n\!\geq\!0$,
Given\,a\,set\,$\mathit{Cls}\!=\!\{C_i \!\mid\! 0\!\leq\!i\!\leq\!n\}$
of\,clauses\,and\,a\,catamorphic\,abstraction\,specification\,$\alpha$,
$\mathit{AddCata}(\mathit{Cls},\alpha)\!=\!\{E_i\!\mid\! 0\!\leq\!i\!\leq\!n$ and $E_i$ is 
obtained from $C_i$ by applying the catamorphism addition rule R3 using $\alpha\}$.
%obtained from $C_i$ by applying the catamor-\\
%\hspace*{35mm}phism addition rule R3 using $\alpha\}$.
%\hspace*{35mm}
%morphism addition rule R3 using  $\alpha\}$.

%%\bcomment{OLD: derived from $C_i$ by
%%performing\\  
%%\hspace*{35mm}***as long as possible*** catamorphism additions using $\alpha\}$.}

\vspace*{-2mm} \noindent \hrulefill

\noindent 
%Given a set $\mathit{Cls}=\{C_1,\ldots,C_n\}$, with $n\!\geq\!0$, 
Given a set $\mathit{Cls}\!=\!\{C_i \!\mid\! 0\!\leq\!i\!\leq\!n\}$
of clauses and a monovariant set $\Delta$ of definitions,\\ 
$\mathit{Fold}(\mathit{Cls}, \Delta)\!=\! \{ E_i\! \mid \!0\!\leq\!i\!\leq\!n$ and $E_i$ is obtained from $C_i$ by 
applying the folding rule~R4 using definitions in $\Delta\}$. %%} % end of comment
%%applying the folding \linebreak 
%%\hspace*{32mm}rule R4 using definitions in $\Delta\}$. %%} % end of comment

%%\bcomment{OLD: it ***as long as\\ 
%%\hspace*{30mm}possible*** using the definitions in $\Delta\}$.}

\vspace*{-2mm} \noindent \hrulefill

Given a set $\mathit{Cls}$ of clauses by applying the erasure addition rule R5, \\
\makebox[27mm][l]{$\mathit{AddErasure}(\mathit{Cls}) =$}
$\{\chi_{\!\mathit{wo}}(C) \mid C$ is a clause in $\mathit{Cls}$ whose head is not \textit{false}$\}$
$\cup$\\
\makebox[27mm][l]{} $\{\chi_{\!\mathit{w\&wo}}(C) \mid C\!\in\! \textit{Cls}\}$.

\vspace*{-2mm} 
\noindent \hrulefill

%%%%%%%%%%%%%%%%%%%%%%%%%%%%
%\vspace*{-2mm}
%\captionsetup[figure]{justification=justified, singlelinecheck=off} %Alberto: not needed
\caption{The $\mathit{Define}$, $\mathit{Unfold}$, $\mathit{AddCata}$, $\mathit{Fold}$, and $\mathit{AddErasure}$ functions. %\protect\footnotemark. 
\label{fig:Functions}}
\end{minipage}
\end{sizepar}
\end{figure}
% -----------------------------------------------------------------------------
% -----------------------------------------------------------------------------

\indent
In the definition of the \textit{Define} function we assume that, for each clause $C$ in \textit{Cls} and each catamorphism atom $\textit{Cata}$ in the body of $C$, there is a program atom $A$ in the body of~$C$ such that $\mathit{adt\mbox{-}vars}(\textit{Cata}) \subseteq \mathit{adt\mbox{-}vars}(A)$. If $A$ is absent for a catamorphism atom having the ADT variable $X$ of sort $\tau$, 
in order to comply with our assumption, we add to the body of $C$ a program atom $\textit{true}_\tau(X)$ that is defined on the (possibly recursive) structure of sort $\tau$ and holds for every $X$ of sort $\tau$. 
 For instance, for the sort $\mathit{list(int)}$, the 
program atom $\textit{true}_\mathit{list(int)}(X)$  
will be defined by the two clauses $\textit{true}_\mathit{list(int)}([\,])$ ~and~ 
$\textit{true}_\mathit{list(int)}([H|T]) \If \textit{true}_\mathit{list(int)}(T)$, where $H$ is an integer 
variable. Note that, by adding to clause $C$ the atom $\textit{true}_\tau(X)$, we get a  clause equivalent to~$C$.

\begin{samepage}
\begin{definition}[Domain of Definitions]\label{def:domain}\nopagebreak
We denote by $\mathcal D$ a maximal set of definitions such that

%(i)  $\mathcal D$ is monovariant,
\noindent\hangindent=8mm
(D1) for every definition $\mathit{newp}(X_1,\ldots,X_k)\leftarrow \textit{Catas}, A$ in $\mathcal D$, 
%where \textit{Catas} is a conjunction of catamorphism atoms and $A$ is a program atom, 
for every ADT variable~$X_i$ occurring %in $\mathit{newp}(X_1,\ldots,X_k)$ and 
in the program atom $A$, 
for each catamorphism predicate \textit{cata} in the conjunction \textit{Catas} of catamorphism atoms, at most
one catamorphism atom of the form $\mathit{cata}(\ldots,X_i,\dots)$ occurs in \textit{Catas}, and  

%%%MAU:
%%%(D1) for every definition $\mathit{newp}(X_1,\ldots,X_k)\leftarrow \textit{Catas}, A$ in $\mathcal D$, 
%%%where \textit{Catas} is a conjunction of catamorphism atoms and $A$ is a program atom, 
%%%for every ADT variable $X_i$ in $A$, and for every catamorphism $c$, there is at most one atom $B$ in \textit{Catas} 
%% ---------
%%\comment{for every ADT variable~$X_i$ occurring in $\mathit{newp}(X_1,\ldots,X_k)$ and in the program atom $A$, 
%%for each catamorphism predicate \textit{cata}, at most
%%one catamorphism atom $\mathit{cata}(\ldots,X_i,\dots)$ occurs in \textit{Catas},} 

%%for every ADT variable $X_i$ in $A$, and for every catamorphism $\mathit{cata}$, there is at most one atom $B$ in \textit{Catas} 
%%with predicate~$c$ and an occurrence of~$X_i$, and   
\noindent
(D2) $\mathcal D$ does not contain %any two 
equivalent definitions.
\end{definition}
\end{samepage}

%a maximal set of definitions where no two elements are equivalent (that is, 
%any monovariant set of definitions containing exactly one representative for each equivalence class of relation $\equiv$. 

It follows directly from our assumptions %the definitions +++ 
that $\mathcal D$ is a finite set. 

Now we define a partial order ($\sqsubseteq$), a join %least upper bound 
operation ($\sqcup$) and a meet %greatest lower bound 
operation ($\sqcap$) %between 
{for definitions and also for monovariant subsets of definitions in~$\mathcal D$.} 

%\vspace*{-1.8mm}
\begin{definition} \label{def:lattice}
Let $D_1$: $\mathit{newp}1(U_1) \If \mathit{Catas}_1, \mathit{Catas}, A$ 
and $D_2$: $\mathit{newp}2(U_2) \If \mathit{Catas}_2, \mathit{Catas}, A$ be two definitions in $\mathcal D$ for the same program atom $A$, where $\mathit{Catas}, \mathit{Catas}_1$, and $\mathit{Catas}_2$
are conjunctions of catamorphism atoms.
We assume that the variables in $D_1$ and $D_2$ have been renamed and the atoms in their bodies have been reordered 
so that $(\mathit{Catas}, A)$ is the maximal common subconjunction of atoms in their bodies, that is,
there exists no atom $\mathit{Cata}$ %$C$ 
in $\mathit{Catas}_1$ and no variant of $D_2$ of the form $\mathit{newp}2(U'_2) \If \mathit{Catas}'_2, \mathit{Catas}, A$, such that $\mathit{Cata}$ %$C$ 
is an atom in $\mathit{Catas}'_2$.
%Let $D_1\mathrm{:}$ $\mathit{newp}1(U_1) \If B_1$ 
%and $D_2\mathrm{:}$ $\mathit{newp}2(U_2) \If B_2$ be two definitions in $\mathcal D$.

%\smallskip
\noindent\hangindent=6mm
(i)~$D_2$ is an \emph{extension} of $D_1$, written $D_1\sqsubseteq D_2$,
if $\mathit{Catas}_1$ is the empty conjunction;%a subconjunction of $\mathit{Catas}_2$.

%\smallskip
\noindent\hangindent=6mm
(ii)~By $D_1 \sqcup D_2$ we denote the definition $D_3$: $\mathit{newp}3(U_3) \If \mathit{Catas}_1, \mathit{Catas}_2, \mathit{Catas}, A$, where
$U_3$ is a tuple consisting of the {distinct} variables occurring in $(U_1,U_2)$;

%\smallskip
\noindent\hangindent=6mm
(iii)~By $D_1 \sqcap D_2$ we denote the definition $D_3$: $\mathit{newp}3(U_3) \If \mathit{Catas}, A$, where
$U_3$ is a tuple consisting of the variables occurring in both $U_1$ and $U_2.$ 

%\smallskip
\noindent\hangindent=6mm
Let $\Delta_1$ and $\Delta_2$ be two monovariant 
subsets of $\mathcal D$.  

\noindent\hangindent=6mm
(iv)~$\Delta_2$ is an \emph{extension} of $\Delta_1$, written $\Delta_1\sqsubseteq \Delta_2$, if 
for each $D_1$ in $\Delta_1$ there exists~$D_2$ in~$\Delta_2$ such that $D_1\sqsubseteq D_2$;

\noindent\hangindent=6mm
(v)~$\Delta_1 \sqcup \Delta_2 = \{D \mid D$ is the only definition in $\Delta_1\cup\Delta_2$ for some program 
atom in 

\hspace{28mm}$\Delta_1\cup\Delta_2 \}\ \cup$ 

\hspace{20mm}$\{D_1\sqcup D_2 \mid D_1$ and $D_2$ are definitions for the same program atom in 

\hspace{38mm}$\Delta_1$ and $\Delta_2$, respectively$\};$

\noindent\hangindent=6mm
(vi)~$\Delta_1 \sqcap \Delta_2  = \{D_1 \sqcap D_2 \mid  D_1$ and $D_2$ are the definitions for the same program atom in 

\hspace{38mm}$\Delta_1$ and $\Delta_2$, respectively$\}.$
\end{definition}

Let $\mathcal P_m(\mathcal D)$ denote the set of monovariant subsets of $\mathcal D$. We have that
$(\mathcal P_m(\mathcal D), \sqsubseteq, \sqcup,  \sqcap)$ is a lattice and, since $\mathcal D$ is a finite set, it is also a complete lattice.
%Now,   % given a set $P$ of CHC clauses, a query~$Q$, and a catamorphic abstraction specification $\alpha$,
We define the operator $\tau_{P\cup\{Q\},\alpha}\!: \mathcal P_m (\mathcal D) \rightarrow \mathcal P_m (\mathcal D)$ as follows:

%%% OLD Not monotone:
%%% \vspace{1mm}
%%$\tau_{P\cup\{Q\},\alpha}(\Delta) = \Bigg\{
%%\begin{matrix*}[l]
%%\mathit{Define}(\mathit{AddCata}(Q,\alpha),\emptyset) & ~\mbox{if}~\Delta\!=\!\emptyset  \\ %[1.mm]
%%\mathit{Define}(\mathit{AddCata}(\mathit{Unfold}(\Delta,P),\alpha),\Delta) & ~\mbox{otherwise}
%%\end{matrix*}$

\smallskip
\indent
$\tau_{P\cup\{Q\},\alpha}(\Delta) =_{\mathit{def}} 
\mathit{Define}\big(\mathit{AddCata}\big(\mathit{Unfold}(\Delta,P) \cup \{Q\},\alpha\big),\Delta\big)$ %~\hfill$(\ddagger)$~~

%% \comment{Il vecchio $\tau$ non era monotono, anche se $\tau_{fix}$ era lo stesso. p. es. 
%% $Q: \leftarrow q(...)$, $\Delta_1 =\emptyset$ e $\Delta_2= \{newp(...) \leftarrow ..., p(...)\}$. 
%% $\tau(\Delta_1)$ ha una nuova def. per $q$, $\tau(\Delta_2)$ no.} 

\smallskip
\noindent
%%\comment{Let us consider the following two cases for the set $\Delta$ of definitions: 
%%(i)~$\Delta\!=\!\emptyset$, and (ii)~$\Delta\!\not=\!\emptyset$.}

%%\comment{Let us consider the following two cases:
%%(i)~$\Delta\!=\!\emptyset$, and (ii)~$\Delta\!\not=\!\emptyset$.
%%
%If $\Delta$ is empty, %If $\Delta\!=\!\emptyset$, 
%%(i)~If $\Delta\!=\!\emptyset$, %since $\mathit{Unfold}(\emptyset,P)\!=\!\emptyset$, we have: 
%% we have that $\tau_{P\cup\{Q\},\alpha}(\emptyset)\!=\!\mathit{Define}(\mathit{AddCata}(\{Q\},\alpha),\emptyset)$. 
%%Thus, $\tau_{P\cup\{Q\},\alpha}(\emptyset)$ is obtained by, first, adding to the body of query~$Q$ suitable catamorphisms 
%%by the function \textit{AddCata}, and then introducing, by the function \textit{Define} (see the (Add) case), a new 
%% definition for each program predicate occurring in~$Q$.}
%%%%thereby deriving a new query $Q'$, and then introducing, by the function \textit{Define} (see the (Add) case), a new 
%%definition for each program predicate occurring in~$Q'$.}
%% % , after adding suitable catamorphisms by the function \textit{AddCata}.
%%
%%(ii)~If $\Delta\!\not=\!\emptyset$, $\tau_{P\cup\{Q\},\alpha}(\Delta)$ is obtained by,
%%first, \comment{adding to the query $Q$ the set of clauses obtained by} unfolding all clauses in $\Delta$ using the  
%%\textit{Unfold} function, then applying the catamorphism addition rule using the \textit{AddCata} function
%%and, finally, applying the \textit{Define} function. 

\clearpage
\noindent
Now, we show that the operator $\tau_{P\cup\{Q\},\alpha}$ is a well defined function from $\mathcal P_m (\mathcal D)$ to itself, that is,
for any $\Delta \in \mathcal P_m (\mathcal D)$, the set $\Delta' = \tau_{P\cup\{Q\},\alpha}(\Delta)$ is an element of $\mathcal P_m (\mathcal D)$.

First, note that: (i)~the \textit{Define} function introduces (see the (Add) case) a new definition for a program predicate only if no definition for that predicate already belongs to $\Delta$,
and (ii)~\textit{Define} replaces (see the (Extend) case) a definition for a program predicate by a new definition for the same predicate. Thus, if $\Delta$ is monovariant, so is $\Delta'$. %is monovariant as well, 
Moreover, no two equivalent clauses will belong to $\Delta'$ (see Point (D2) of Definition~\ref{def:domain}).

Note also that, due to the definition of function \textit{AddCata} (see, in particular, Point~(ii) of Rule R3 applied by that function), Point (D1) of Definition~\ref{def:domain} holds,
and in particular, for every ADT variable $X_i$ in the body of any new definition in $\Delta'$, and for every catamorphism predicate~$cata$, %there is at most one atom with predicate $c$ and an occurrence of~$X_i$.}
there is at most one catamorphism atom of the form $\mathit{cata}(\sdots,X_i,\sdots)$.

\medskip
\noindent
\begin{minipage}{1\textwidth}
\begin{lemma}[Existence and Uniqueness of the Fixpoint of $\tau_{P\cup\{Q\},\alpha}$]  \label{lemma:fixpoint}
The operator $\tau_{P\cup\{Q\},\alpha}$ is monotonic on the finite lattice $\mathcal P_m(\mathcal D)$.
Thus, it has a least fixpoint $\textit{lfp}(\tau_{P\cup\{Q\},\alpha})$, also denoted $\tau_{\mathit{fix}}$, 
which is equal to $\tau_{P\cup\{Q\},\alpha}^n(\emptyset)$,  for some natural number $n$.
%$\textit{lfp}(\tau_{P\cup\{Q\},\alpha})=\tau_{P\cup\{Q\},\alpha}^n(\emptyset)$,  for some $n\geq 0$.
\end{lemma}
\end{minipage}

\begin{proof}
In order to prove the monotonicity of $\tau_{P\cup\{Q\},\alpha}$, let us 
assume that $\Delta_1$ and $\Delta_2$ are two sets of monovariant definitions in 
$\mathcal P_m (\mathcal D)$, with $\Delta_1 \sqsubseteq \Delta_2$. 
Let $D_1\in \tau_{P\cup\{Q\},\alpha}(\Delta_1)$ be a definition for program atom $A$. 
We consider two cases.

\smallskip
\noindent
(\textit{Case} 1) There is no definition for $A$ in $\Delta _1$. Then, by construction, 
according to the \textit{Define} function (see Figure~\ref{fig:Functions}),
%%the inner for-loop \comment{Non solo inner, anche outer: lo stesso A puo' occorrere in diverse clausole di Cls}
%%of the \textit{Define} function (see Figure~\ref{fig:Functions}),
$D_1$ can be viewed as the result of a sequence of join operations of the form: $E_0\sqcup E_1 \sqcup \sdots \sqcup E_n$, 
with $n\!\geq\!0$, where: 
%(we assume associativity from left to right). In that sequence, 
(1)~clause~$E_0$ has been obtained by the (Add) case of \textit{Define}, and (2)~for $i\!=\!1,\sdots, n$, clause 
$E_0\sqcup\sdots\sqcup E_{i}$ is a clause
obtained by the (Extend) case of \textit{Define} from clause $E_0\sqcup\sdots\sqcup E_{i-1}$.
In particular, for all $i\!=\!0,\sdots,n$, clause~$E_i$ 
is a clause of the form $\mathit{newp}_i(V_i) \If \mathit{Catas_i}, A$ obtained from a clause $H\leftarrow c, G$ 
(here and below in this proof $H$ may be \textit{false}) 
in $\mathit{AddCata}(\mathit{Unfold}(\Delta_1,P) \cup \{Q\},\alpha)$ such that~$A$ is a program 
atom in $G$ and $\mathit{Catas_i}$ is the conjunction of all catamorphism atoms~$F$ in~$G$ 
with $\mathit{adt\mbox{-}vars}(F)\subseteq\mathit{adt\mbox{-}vars}(A)$.  

%%
%%\bcomment{OLD:
%%
%%$D_1 = ((E_0\sqcup E_1)\sqcup\sdots\sqcup E_n)$, where $E_i$ is a clause of the form $\mathit{newp}_i(V_i) \If  \mathit{Catas_i}, A$ obtained from a 
%%clause $H\leftarrow c, G$ \comment{(where $H$ is possibly \textit{false})} in 
%%$\mathit{AddCata}(\mathit{Unfold}(\Delta_1,P)\comment{\cup \{Q\}},\alpha)$ 
%%such that $A$ is a program atom in $G$ and $\mathit{Catas_i}$ is the conjunction of all catamorphism atoms $F$ in $G$ with $\mathit{adt\mbox{-}vars}(F)
%%\subseteq\mathit{adt\mbox{-}vars}(A)$.
%%Indeed, clause~$E_0$ has been obtained by the (Add) case of \textit{Define} and, for $i=1,\ldots, n$, $E_0\sqcup\ldots\sqcup E_i$ is a clause 
%%\textit{ExtD} computed by the (Extend) case of \textit{Define}.}
%%
%%+++

\smallskip
\noindent
(\textit{Case} 2) There is a definition $E_0$ for $A$ in $\Delta _1$. Then, similarly to Case 1,  by construction, 
$D_1 = E_0\sqcup\sdots\sqcup E_n$, where, for $i\!=\!1,\sdots, n$, with $n\!\geq\!0$, $E_0\sqcup\sdots\sqcup E_i$ is a clause obtained 
by the (Extend) case of \textit{Define}.

\smallskip
Now, since $\Delta_1 \sqsubseteq \Delta_2$, for each clause  $H\leftarrow c, G$ 
%\bcomment{(where $H$ is possibly \textit{false})}
in $\mathit{AddCata}(\mathit{Unfold}(\Delta_1,P)\cup \{Q\},\alpha)$, there exists a clause $H\leftarrow c, C, G$ 
%(where $H$ is possibly \textit{false}) 
in the set of clauses $\mathit{AddCata}(\mathit{Unfold}(\Delta_2,P)
\cup \{Q\},\alpha)$, where $C$ is a conjunction of catamorphism atoms, and then, by construction,\linebreak $\textit{Define}
(\mathit{AddCata}(\mathit{Unfold}(\Delta_2,P)\cup \{Q\},\alpha),\Delta_2)$ contains, for $i\!=\!1,\sdots,n$, a 
clause $E'_i$, with $E_i\sqsubseteq E'_i$.
Then, there exists $D_2 \in \tau_{P\cup\{Q\},\alpha}(\Delta_2)$ such that $D_1\! =\! (E_0\sqcup\sdots\sqcup E_n) \sqsubseteq  
(E'_0\sqcup\sdots\sqcup E'_n) \sqsubseteq 
(E'_0\sqcup\sdots\sqcup E'_n \sqcup F_{1}\sqcup\sdots\sqcup F_r) = D_2$, with $r\geq 0$.
(Note that, since  $\Delta_1\sqsubseteq \Delta_2$, in $\mathit{AddCata}(\mathit{Unfold}(\Delta_2,P) \cup \{Q\},\alpha)$ there may be clauses
that are derived from definitions in $\Delta _2$ that are not extensions of definitions in $\Delta_1$. 
In the bodies of those clauses there may be some variants of $A$ that determine $r$ extra applications of the (Extend) case of \textit{Define}.)
Therefore, by Definition~\ref{def:lattice}, $\tau_{P\cup\{Q\},\alpha}(\Delta_1) \sqsubseteq \tau_{P\cup\{Q\},\alpha}(\Delta_2)$. 

Thus, $\tau_{P\cup\{Q\},\alpha}$ is monotonic with respect to $\sqsubseteq$.
Since $\mathcal P_m(\mathcal D)$ is a finite, hence complete, lattice, $\tau_{P\cup\{Q\},\alpha}$ 
has a  least fixpoint $\textit{lfp}(\tau_{P\cup\{Q\},\alpha})$, which can be computed as 
$\tau_{P\cup\{Q\},\alpha}^n(\emptyset)$, for some natural number $n$.
\hfill\end{proof}

%%% Maurizio 20.03.24
%%%\comment{Confermo $\sqsubseteq D_2$, non $= D_2$, vedi esempio seguente}
%%%
%%%\begin{verbatim}
%%%Delta1: 
%%%E0: new1 :- c1, p
%%%
%%%Delta2: 
%%%E0': new2 :- c1, c2, p
%%%     new3 :- c3, q
%%%        
%%%p :- p.
%%%q :- p.
%%%
%%%alpha: p ==> c1
%%%       q ==> c1
%%%
%%%Unfold;AddCata Delta1 
%%%
%%%E1: new1 :- c1, p
%%%
%%%Define:
%%%D1 = E0 U E1 = E0
%%%
%%%Unfold;AddCata Delta2
%%%
%%%E1': new2 :- c1, c2, p
%%%E2': new3 :- c1, c3, p
%%%
%%%Define:
%%%D2: = E0' U E1' U E2' = new4 :- c1, c2, c3, p
%%%                        new3 :- c3, q
%%%\end{verbatim}

%Let $\tau_{\mathit{fix}}$ denote the set $\textit{lfp}(\tau_{P\cup\{Q\},\alpha})$ of definitions.  
%By construction, $\tau_{\mathit{fix}}$ is monovariant.
%
Now, we define our transformation algorithm \abs\  as follows:

\smallskip

\abs$(P\cup\{Q\},\alpha) = \mathit{AddErasure}\big(\mathit{Fold}\big(\mathit{AddCata}\big(\mathit{Unfold}(\tau_{\mathit{fix}},P)
\cup \{Q\},\,\alpha\big),\tau_{\mathit{fix}}\big) \big)$

\smallskip

%Then, starting from a set $P$ of CHC clauses, by using the set $\tau_{\mathit{fix}}$
%$\textit{lfp}(\tau_{P\cup\{Q\},\alpha})$ 
%of definitions, the transformation algorithm 

%%\noindent
%%\comment{In words, we have that, first, \abs\ 
%%applies the {\it Unfold} function to the set $\tau_{\mathit{fix}}$ of definitions. Then, it adds to the 
%%derived set the query $Q$.  Next, it adds catamorphisms using the {\it AddCata} function based on the catamorphic abstraction 
%%$\alpha$, and it applies the {\it Fold} function to the set of clauses 
%%obtained so far, using the definitions in $\tau_{\mathit{fix}}$. %derived by unfolding and catamorphism addition. 
%%Finally, \abs\  applies %the erasure addition rule~R5 according to 
%%the \textit{AddErasure} function.}

%\comment{Alb: $\mathit{Strengthen}\big(\mathit{Unfold}(\Delta,P)\,Q\big)$: questa parte e' comune al calcolo del fixpoint delle definizioni. Forse si puo' scrivere questa parte una volta sola.}

\noindent
The termination of \abs\ follows immediately from
the fact that %each of 
the functions \textit{Unfold}, \textit{AddCata}, \textit{Fold}, and \textit{AddErasure} terminate and
the least fixpoint $\tau_{\mathit{fix}}$
%$\textit{lfp}(\tau_{P\cup\{Q\},\alpha})$ 
is computed in a finite number of steps (see Lemma~\ref{lemma:fixpoint}). Thus, by the correctness of
the transformation rules (see Theorem~\ref{thm:Corr}), we 
get the following result.

%\vspace*{-2mm}
\begin{theorem}[Total Correctness of {Algorithm}~\abs]  
\label{thm:termination}
\abs~terminates for any set $\mathit{P}$ of definite clauses, query $Q$, and catamorphic abstraction specification~$\alpha$.
Also, $P \cup \{Q\}$ is satisfiable if and only if
\abs$(P\cup \{Q\},\alpha)$ is satisfiable.
\end{theorem}

%\vspace*{-1mm} 
Finally, we would like to comment on the fact that our transformation algorithm~\abs~introduces a monovariant set of 
definitions. Other definition introduction policies could have been considered. In particular, one could introduce more than 
one definition for each program predicate, thus producing a \textit{polyvariant} set of definitions. The choice between 
monovariant and polyvariant sets of definitions %transformations 
has been subject to ample discussion in the literature \cite{DeAngelisFGHPP21} and both have advantages and disadvantages.  
We will show in the next section that our technique %is adequate and 
performs quite well in our benchmark. However, we leave a more accurate 
experimental evaluation to future work.
%\vspace*{-4mm}

\section{Implementation and Experimental Evaluation} 
\label{sec:Experiments}
%\newcommand{\isasorted}{{\it is\us asorted}}

%Eldarica~\cite{HoR18} 
%(with the built-in catamorphism \textit{size}) and Spacer~\cite{Ko&16}.

%This section is divided into two parts: (i)~Section~\ref{subsec:implementation}, in which we will 
%indicate how we have implemented 
%the algorithm~\abs\ presented in Section~\ref{subsec:Strategy},
%and (ii)~Section~\ref{subsec:experiments}, in which we will evaluate the results obtained 
%in our experiments.

%\comment{2 sottosezioni? Implementazione e Esperimenti}

%\subsection{Implementation of Algorithm \abs} 
%\label{subsec:implementation}

In this section we provide some details on the implementation of algorithm~\abs,
and on its experimental evaluation.

\bigskip

\noindent
\textbf{Implementation.}
We have implemented algorithm~\abs\ in a tool, called \vericatabs, 
based on VeriMAP \cite{De&14b}, which is a system 
for transforming CHCs.
In order to check satisfiability of sets of CHCs (before and after
their transformation) we have used the following two solvers: (i)~Eldarica  
(v.\,2.0.9)~\cite{HoR18}, 
and (ii)~Z3~(v.\,4.12.2) \cite{DeB08} with the SPACER engine \cite{Ko&16} 
and the {\it global guidance} option~\cite{ KrishnanCSG20}.

The tool \vericatabs manipulates clauses as indicated in the following three phases.

\smallskip
\noindent
(\textit{Phase}~1) A pre-processing phase. In this phase 
\vericatabs produces a catamorphic abstraction specification~$\alpha$
starting from: (i)~a given set $P$ of CHCs, and (ii)~the catamorphic abstractions for the ADTs occurring in~$P$.
For instance, in the case of our introductory example {\it{double}} (see Figure~\ref{fig:ProgCataQuery}), 
%in Section~\ref{sec:IntroExample},
Phase~1 produces the catamorphic abstraction specifications for 
$\mathit{double}$, $\mathit{eq}$, and $\mathit{append}$ we have 
listed in Example~\ref{ex:double}, starting from clauses~1--6 % 1--4
and the catamorphic abstraction
$\mathit{cata}_{\mathit{list}(\mathit{int})} =_\mathit{def} {\mathit{listcount}(X,L,N)}$,
%of Example~\ref{ex:double},
%%for the ADT sort 
%%$\mathit{list}(\mathit{int})$ shown in Example~\ref{ex:double} of Section~\ref{subsec:CataAbsSpecif}, 

In the following example, referring to a treesort algorithm, %~\ref{ex:treesortas}, 
we present the \vericatabs syntax for representing: (i)~the catamorphic abstractions given in input,
 using the directive \verb|cata_abs|, 
and (ii)~the catamorphic abstraction specifications produced in output, after Phase~1, using 
the directive~\verb|spec|. 
%and the output produced by the pre-processing phase for a set of CHCs encoding the

%%that acts on the ADT sorts $\mathit{list(int)}$ and $\mathit{tree(int)}$ and uses the catamorphisms
%%\textit{listcount} and \textit{trecount} and also the predicate {\it visit} and
%%{\it treesort}.

\begin{example}\label{ex:treesortas}
Let \verb|treesort(L,S)| and \verb|visit(T,L)| be two atoms included in a CHC encoding of the treesort algorithm.
The atom \verb|treesort(L,S)| holds if and only if \verb|S| is the list of integers obtained by applying the treesort algorithm to the list \verb|L| of integers. The auxiliary % ancillary 
atom \verb|visit(T,L)| holds if and only if \verb|L| is the list of integers obtained by a depth first visit of the tree \verb|T| with integers at its nodes. %, starting from its leftmost node. 
The catamorphic abstractions for the ADT sorts $\mathit{list(int)}$ and $\mathit{tree(int)}$ 
used by our tool \vericatabs during Phase~1, are as follows:
%encoded through the use of~\verb|cata_abs| as follows:

\vspace{1mm}
  \verb|:- cata_abs list(int) ==> listcount(X,L,C).| 
  
  \verb|:- cata_abs tree(int) ==> treecount(X,T,C).|

\vspace{1mm}
\noindent
The catamorphisms  \verb|listcount(X,L,B)| and \verb|treecount(X,T,A)| count the occurrences of the integer \verb|X|  in the list \verb|L|
and in the tree \verb|T|, respectively.
In general, the directive \verb|cata_abs| for a sort $\tau$ is as follows:

\vspace{1mm}
  \verb|:- cata_abs | $\tau$  \verb| ==> | catamorphisms acting on~$\tau$\verb|.|

\vspace{1mm}

%%To the left of\, \verb|==>|, \verb|cata_abs| specifies the sort $\tau$, and to the right of\, 
%%\verb|==>|, \verb|cata_abs|  specifies the
%%catamorphisms
%%% (or, in general, the conjunction of distinct catamorphisms) 
%%acting on~$\tau$.
%
For the program predicates \verb|treesort| and \verb|visit|,  the catamorphic abstraction specifications 
produced by \vericatabs after Phase~1,
are as follows: % encoded through~\verb|spec| as follows:

\vspace{1mm}
  \verb|:- spec treesort(L,S) ==> X=Y, listcount(X,S,A), listcount(Y,L,B).|\nopagebreak
  
  \verb|:- spec visit(T,L) ==> X=Y, treecount(X,T,A), listcount(Y,L,B).|

\vspace{1mm}
\noindent
Note that both the tree catamorphism \verb|treecount(X,T,A)| and the list  catamorphism \verb|listcount(Y,L,B)| occur in the catamorphic specification for \verb|visit(T,L)|.~\hfill$\Box$
\end{example}

\smallskip
\noindent
(\textit{Phase}~2) A fold/unfold transformation phase. 
In this phase \vericatabs computes the fixpoint $\tau_{\mathit{fix}}$ and the set $T_w$ of clauses,
 which is 
$\mathit{Fold}\big(\mathit{AddCata}\big(\mathit{Unfold}(\tau_{\mathit{fix}},P) \cup \{Q\},\,\alpha\big),\tau_{\mathit{fix}})$. For the {\it{double}} introductory example (see Figure~\ref{fig:ProgCataQuery}), we have that $P$ is 
the set $\{1,\ldots,6\}$ of clauses, query~$Q$ is clause~7, and $\alpha$ is the set of catamorphic abstraction
specifications produced at Phase~1 (see Example~\ref{ex:double}).
Now, $\tau_{\mathit{fix}}$ is the set  $\{D1,D2,D3,D4\}$ of definitions  listed in Figure~\ref{fig:TransfProgDefs} and the set~$T_w$ 
is as follows: % (modulo variable renaming):

\vspace*{1mm}
$\mathit{false} \leftarrow C\!=\!2D+1,\ \mathit{new}1(A,E,F,G,C)$

$\mathit{new}1(A,B,C,E,F) \leftarrow \mathit{new}2(A,M,K,E,F,B,C),\ \mathit{new}3(A,M,K,B,C)$
      
$\mathit{new}2(A,B,C,B,C,[\,],G) \leftarrow G\!=\!0,\ \mathit{new}4(A,B,C)$\nopagebreak

$\mathit{new}2(A,B,C,[E|F],G,[E|J],K) \leftarrow G\!=\!\mathit{ite}(A\!=\!E,N\!+\!1,N),\ K\!=\!\mathit{ite}(A\!=\!E,P\!+\!1,P),$\nopagebreak

\hspace{20mm}
$\mathit{new}2(A,B,C,F,N,J,P)$\nopagebreak
      
$\mathit{new}3(A,B,C,B,C) \leftarrow \mathit{new}4(A,B,C)$  

$\mathit{new}4(A,[\,],B) \leftarrow  B\!=\!0$

$\mathit{new}4(A,[B|C],D) \leftarrow  D\!=\!\mathit{ite}(A\!=\!B,F\!+\!1,F),\ \mathit{new}4(A,C,F)$

\vspace{1mm}
\smallskip
\noindent
(\textit{Phase}~3) A post-processing phase. In this phase, 
\vericatabs produces the following two additional sets of clauses by applying 
the $\mathit{AddErasure}$ function to~$T_w$:

\noindent\hangindent=6mm
(i) $T_{\mathit{wo}}=\{\chi_{\mathit{wo}}(C) \mid C$ is a clause in $T_w\}$, that is, $T_{\mathit{wo}}$ is made out of the clauses in $T_w$ where every atom with 
ADT arguments has been replaced by its corresponding atom
without ADT arguments, and

\noindent\hangindent=6mm
(ii) $T_{\mathit{w\&wo}}=\{\chi_{\mathit{w\&wo}}(C) \mid C$ is a clause in $T_w\} \cup \overline{T}_{\mathit{wo}}$,
that is, $T_{\mathit{w\&wo}}$ is made out of the clauses in {\it either\/} (ii.1)~$T_w$, where every 
atom \textit{in the body\/} with ADT arguments is paired with its corresponding atom without ADT arguments,
{\it or} (ii.2)~$\overline{T}_{\!\mathit{wo}} = \{\chi_{\mathit{wo}}(C) \mid C$ is a clause in $T_\mathit{w}$ whose head is not \textit{false}\}.

\smallskip
\noindent
$T_{\mathit{w\&wo}}$ is, indeed, the set of clauses computed by our transformation algorithm \abs.
The other two sets $T_{\mathit{w}}$ and $T_{\mathit{wo}}$, produced by \vericatabs, will be used for comparing
and analysing the features of $T_{\mathit{w\&wo}}$, as we do in the experimental evaluation below. 
%Table~\ref{tab:exper}.
 
\smallskip
For our introductory example {\it{double}} (see Figure~\ref{fig:ProgCataQuery}), at the end of Phase~3,
\vericatabs produces the following two sets of clauses (clause numbers refer to Figure~\ref{fig:TransfProgDefs}):

%where the clause numbers refer to Figure~\ref{fig:TransfProgDefs}:
%where $\overline{T}_{\!\mathit{wo}}\!=\! \{18,19,20,21,22,23\}$ (see Figure~\ref{fig:TransfProgDefs}): 

%$\overline{T}_{\!\mathit{wo}}$ denotes clauses 18--23 (see Figure~\ref{fig:TransfProgDefs}): 
\vspace*{.5mm}
\hangindent=15mm
${T}_{\mathit{wo}}\!=\! \{\mathit{false} \leftarrow C\!=\!2D\!+\!1,\ {\newiwoADTs}(A,\!F,\!C)\}$ 
$ \cup \ \{18,\ldots, 23\}$, and 
%% (these clauses make out the set $\overline{{T}}_{\mathit{wo}}$), and
%%
%%clauses 18--23 that co the set $\overline{T}_{\!\mathit{wo}}$, and

%%
%%${T}_{\mathit{wo}}= \{\mathit{false} \leftarrow C\!=\!2D+1, {\newiwoADTs}(A,F,C)\}\,\cup\,\{18,19,20,21,22,23\}$,
%%\ and
%%\hspace*{30mm}where $\overline{T}_{\!\mathit{wo}}= \{18,19,20,21,22,23\}$, and 
\vspace*{.5mm}

%%$T_{\mathit{w\&wo}}=\{11,12,13,14,15,16,17,1\}$.

%%$T_{\mathit{w\&wo}}=\{11, 12, 13,$ $14, 15, 16, 17\}
%%\,\cup\, \overline{T}_{\!\mathit{wo}}$. % where $\overline{T}_{\!\mathit{wo}}= \{18,19,20,21,22,23\}$. 

\indent
\hangindent=4mm
$T_{\mathit{w\&wo}} = \{11,\ldots,23\}$. %, which is made out of clauses 11--23. 

\vspace*{.5mm}
\noindent
The set $\{18,\ldots,23\}$ of clauses is $\overline{{T}}_{\mathit{wo}}$ of Point~(ii.2) above.
%%
%%\vspace*{.5mm}
%%\noindent

%\subsection{Experimental Evaluation} 
%\label{subsec:experiments}

\bigskip

\noindent
\textbf{Experimental Evaluation.} 
Our benchmark consists of 228 
{sets of CHCs that encode properties of} various sorting algorithms 
(such as bubblesort, heapsort, insertionsort, mergesort, quicksort, 
selectionsort, and treesort), and simple list and tree manipulation 
algorithms (such as appending and reversing lists, constructing permutations, 
deleting copies of elements, manipulating binary search trees). 
Properties of those algorithms are expressed via 
catamorphisms. Here is a non-exhaustive list of the
catamorphisms we used:
(i)~$\mathit{size(L,S)}$, (ii)~$\mathit{listmin(L,Min)}$, 
(iii)~$\mathit{listmax(L,Max)}$, and 
(iv)~$\mathit{sum(L,Sum)}$ computing, respectively, the size~$S$ of list~$L$, 
the minimum~{\it Min}, the maximum~{\it Max}, and 
the sum {\it Sum} of the elements of 
list~$L$, (v)~\isasorted\/$(L,\mathit{BL})$, which holds with 
$\mathit{BL\!=\!true}$ if and only if list~$L$ is ordered in weakly ascending order, 
(vi)~{\it allpos\/}$(L,B)$, which holds with
$\mathit{B\!=\!true}$ if and only if list~$L$ is made out of all positive elements, 
(vii)~$\mathit{member}(X,L,B)$, which holds with
$\mathit{B\!=\!true}$ if and only if $X$ is an element of the list~$L$, and 
(viii)~$\mathit{listcount}(X,L,N)$, which holds if and only if~$N$ is 
the number~$(\geq\!0)$ of occurrences of~$X$ in the list $L$. 
For some properties, we have used
more than one catamorphism at a time 
and, in particular, for lists of integers, we have used 
the conjunction of $\mathit{member}$ and
$\mathit{listcount}$, and for different properties, we have also used 
the conjunction of $\mathit{listmin, 
listmax}$, and $\mathit{\isasorted}$, as already indicated in the paper. 

A property holds if and only if its CHC encoding via a query $Q$ is 
satisfiable, and a verification task consists in using a CHC solver to check 
the satisfiability of~$Q$.
When the given property holds for a set $P$ of clauses, the {solver} should 
return {\it sat\/} %\comment{($s$ in Table~\ref{tab:exper})} 
and the property is said 
to be a {\it sat} property. Analogously, when a property does not hold,
the {solver} should return {\it unsat\/} %\comment{($u$ in Table~\ref{tab:exper})}
and the property is said 
to be an {\it unsat} property. 
In our benchmark, for each verification task of a {\it sat} property,  
we have considered  a companion verification task
whose CHCs have been modified 
so that the associated property is {\it{unsat}}. In particular, we have 114 
{\it sat} properties and
114 {\it unsat} properties.

We have performed our experiments on an Intel(R) Xeon(R) Gold 6238R 
CPU 2.20GHz with 221GB RAM under CentOS 
with a timeout of 600s per verification task. 
The results of our experiments are reported in Table~\ref{tab:exper}.
The \vericatabs tool and the benchmarks are available at \url{https://fmlab.unich.it/vericatabs}. 
%%%%%%%
% https://www.tablesgenerator.com/latex_tables

\begin{table}[htbp]
% old: (input: $T_{\mathit{w\&wo}}$; ${\mathit{w\amp wo}}$-columns) 
\caption{Properties proved by the solvers Eldarica and Z3 %~{\rm{(SPACER)}}
{\it{before}}  and {\it{after}} 
the transformation performed by algorithm~\abs. 
In the {\it{before}} case, the input to the solver is the source set of clauses (\mbox{{\it{src}}-columns}), and 
in the {\it{after}} case, the input is $T_{\mathit{w\&wo}}$
\mbox{($T_{\mathit{w\amps wo}}$-columns)}. 
The columns occur in pairs referring to the \textit{sat} properties \mbox{($s$-columns)} and the \textit{unsat} properties
\mbox{($u$-columns)}, respectively. 
The two $T_w$-columns and the two $T_\mathit{wo}$-columns refer to the input $T_{\mathit{w}}$ and~$T_{\mathit{wo}}$, respectively.
%% Similarly, for Z3, %~{\rm{(SPACER)}}, 
%%instead of Eldarica.
The last column shows the time (in seconds) taken by~\abs\ as implemented by \vericatabs.}
%%\comment{alb: manca il tempo totale di prova = TimeTransformation + TimeSolver. Mettere nel testo
%%un paragrafo per dire in media quanto tempo impiega la prova di una propriet\`a che non va in timeout
%%e quante vanno in TO.}
\begin{center}
\begin{tabular}{|@{\hspace{0.5mm}}l@{\hspace{0.5mm}}@{\hspace{0mm}}r@{\hspace{0.5mm}}|@{\hspace{0mm}}r@{\hspace{0.5mm}}||@{\hspace{0mm}}r@{\hspace{0.5mm}}|@{\hspace{0mm}}r@{\hspace{0.5mm}}|@{\hspace{0mm}}r@{\hspace{0.5mm}}
|@{\hspace{0mm}}r@{\hspace{0.5mm}}|@{\hspace{0mm}}r@{\hspace{0.5mm}}|@{\hspace{0mm}}r@{\hspace{0.5mm}}|@{\hspace{0mm}}r@{\hspace{0.5mm}}|@{\hspace{0mm}}r@{\hspace{0.5mm}}||
@{\hspace{0mm}}r@{\hspace{0.5mm}}|@{\hspace{0mm}}r@{\hspace{0.5mm}}|@{\hspace{0mm}}r@{\hspace{0.5mm}}|@{\hspace{0mm}}r@{\hspace{0.5mm}}|@{\hspace{0mm}}r@{\hspace{0.5mm}}|@{\hspace{0mm}}r@{\hspace{0.5mm}}|@{\hspace{0mm}}r@{\hspace{0.5mm}}|@{\hspace{0mm}}r@{\hspace{0.5mm}}||@{\hspace{0mm}}r@{\hspace{0.5mm}}|}
\hline \textrm{~}  &  \multicolumn{2}{@{\hspace{1mm}}c||}{\textit{}}  
%\hline \textrm{~}  &  \multicolumn{2}{@{\hspace{1mm}}c||}{\textit{\!Proper-}}  
& \multicolumn{8}{c||}{\textrm{Eldarica}}  & \multicolumn{8}{c||}{\textrm{Z3}} & \textrm{\footnotesize{\,Transf}}  \\
\cline{4-19}
\multicolumn{3}{|r||}{\textit{Properties}} 
%% & \multicolumn{2}{c||}{\textit{erties}} 
& \multicolumn{2}{c|}{$\mathit{src}$} & \multicolumn{2}{c|}{$T_\mathit{w\amps wo}$} & \multicolumn{2}{c|}{$T_\mathit{w}$} & 
\multicolumn{2}{c||}{$T_\mathit{wo}$} & \multicolumn{2}{c|}{$\mathit{src}$} & \multicolumn{2}{c|}{$T_\mathit{w\amps wo}$} & 
\multicolumn{2}{c|}{$T_\mathit{w}$} & \multicolumn{2}{c||}{$T_\mathit{wo}$} & \textrm{\footnotesize{time}~}    \\ 
%%%old:
%%& \multicolumn{2}{c|}{$\mathit{src}$} & \multicolumn{2}{c|}{$\mathit{w\amp wo}$} & \multicolumn{2}{c|}{$\mathit{w}$} & 
%%\multicolumn{2}{c||}{$\mathit{wo}$} & \multicolumn{2}{c|}{$\mathit{src}$} & \multicolumn{2}{c|}{$\mathit{w\amp wo}$} & 
%%\multicolumn{2}{c|}{$\mathit{w}$} & \multicolumn{2}{c||}{$\mathit{wo}$} & \textrm{\footnotesize{time}~}    \\ 

%\multicolumn{1}{|l|}{\textrm{Program}} &  \textit{sat} & \textit{uns} &  \textit{sat}   & \textit{uns}  & \textit{sat}   & \textit{uns}  & \textit{sat}  & \textit{uns} & \textit{sat}  & \textit{uns}  & \textit{sat}   & \textit{uns}  & \textit{sat}   & \textit{uns}  & \textit{sat}  & \textit{uns} & \textit{sat}  & \textit{uns}  & $T$~~ \\
\cline{4-19}
\multicolumn{1}{|l|}{\textrm{Programs}} &  \textit{s} & \textit{u} &  \textit{s}   & \textit{u}  & \textit{s}   & \textit{u}  & \textit{s}  & \textit{u} & \textit{s}  & \textit{u}  & \textit{s}   & \textit{u}  & \textit{s}   & \textit{u}  & \textit{s}  & \textit{u} & \textit{s}  & \textit{u}  & $T$~~ \\
\hline
\multicolumn{1}{|l|}{\textrm{Append}}
%\textrm{Append}    
& 4      & 4   & 0   & 3    & 3   & 3    & 3  & 4   & 3  & 4    & 0   & 4    & 4   & 4    & 4  & 4   & 4  & 4    & 6.4  \\
\multicolumn{1}{|l|}{\textrm{Bubblesort}}
%\textrm{Bubblesort}
& 9 & 9   & 2   & 9    & 9   & 9    & 9  & 9   & 9  & 9    & 0   & 9    & 9   & 9    & 9  & 9   & 9  & 9    & 15.8 \\
\multicolumn{1}{|l|}{\textrm{BinSearchTree}}
%\textrm{BinSearchTree} 
& 8 & 8   & 0   & 7    & 0   & 5    & 2  & 7   & 4  & 8   & 0   & 8    & 8   & 8    & 7  & 8   & 8  & 8    & 19.2 \\
\multicolumn{1}{|l|}{\textrm{DeleteCopies}}
%\textrm{DeleteCopies} 
& 7& 7  & 0   & 7    & 4   & 7    & 3  & 7   & 6  & 7    & 0   & 7    & 7   & 7    & 7  & 7   & 7  & 7    & 11.1 \\ %\hline
\multicolumn{1}{|l|}{\textrm{Heapsort}}
%\textrm{Heapsort}
& 7 & 7   & 0   & 7    & 2  & 7   & 0 & 7  & 4 & 7   & 0  & 7   & 7  & 7   & 3 & 7   & 5  & 7    & 13.5 \\ 
\multicolumn{1}{|l|}{\textrm{Insertionsort}}
%\textrm{Insertionsort}
& 9 & 9   & 2   & 9    & 9   & 9    & 9  & 9   & 9  & 9    & 0   & 9    & 9   & 9    & 9  & 9   & 9  & 9    & 16.0 \\
\multicolumn{1}{|l|}{\textrm{Member}}
%\textrm{Member}   
& 1 & 1   & 0   & 1    & 1   & 1    & 1  & 1   & 1  & 1    & 0   & 1    & 1   & 1    & 1  & 1   & 1  & 1    & 1.7  \\
\multicolumn{1}{|l|}{\textrm{Mergesort}}
%\textrm{Mergesort}
& 9 & 9   & 0   & 9    & 1   & 9    & 2  & 9   & 4  & 9    & 0   & 9    & 9   & 9    & 3  & 9   & 7  & 9    & 14.1 \\ %\hline
\multicolumn{1}{|l|}{\textrm{Permutations}}
%\textrm{Permutations}
   & 7 & 7   & 2   & 7    & 7   & 7    & 7  & 7   & 7  & 7    & 0   & 7    & 7   & 7    & 7  & 7   & 7  & 7    & 12.4 \\ 
\multicolumn{1}{|l|}{\textrm{QuicksortA}}
%\textrm{QuicksortA} 
   & 8 & 8   & 0   & 6    & 2   & 3    & 1  & 6   & 5  & 8    & 0   & 8    & 8   & 8    & 8  & 8   & 8  & 8    & 14.3 \\
\multicolumn{1}{|l|}{\textrm{QuicksortC}}
%\textrm{QuicksortC}
    & 8 & 8   & 0   & 8    & 1   & 7    & 1  & 8   & 3  & 8    & 0   & 8    & 6   & 8    & 5  & 8   & 6  & 8    & 13.4 \\
\multicolumn{1}{|l|}{\textrm{Reverse}}
%\textrm{Reverse} 
   & 12& 12  & 1   & 12   & 6   & 11   & 6  & 12  & 11 & 12   & 0   & 12   & 11  & 12   & 3  & 12  & 11 & 12   & 20.9 \\ %\hline
\multicolumn{1}{|l|}{\textrm{ReverseAcc}}
%\textrm{ReverseAcc} 
  & 8 & 8   & 0   & 8    & 6   & 7    & 7  & 8   & 7  & 8    & 0   & 8    & 8   & 8    & 8  & 8   & 7  & 8    & 15.6 \\ 
\multicolumn{1}{|l|}{\textrm{ReverseRev}}
%\textrm{ReverseRev}
 & 2 & 2   & 0   & 2    & 0   & 0    & 0  & 0   & 2  & 2    & 0   & 2    & 2   & 2    & 0  & 2   & 2  & 2    & 3.8   \\
\multicolumn{1}{|l|}{\textrm{Selectsort}}
%\textrm{Selectsort}
& 9 & 9   & 2   & 9    & 7   & 8    & 7  & 9   & 8  & 9    & 0   & 9    & 8   & 9    & 8  & 9   & 8  & 9  & 14.2      \\
\multicolumn{1}{|l|}{\textrm{Treesort}}
%\textrm{Treesort} 
   & 6    & 6      & 0 & 6  & 1 & 6  & 1       & 6 & 4       & 6  & 0 & 6  & 5 & 6  & 1       & 6 & 5       & 6  & 10.2 \\
\hline
\multicolumn{3}{|c||}{} & $\,E_1$ & $E_2$  & $\hspace*{.7mm}E_3$\hspace*{.4mm}   & $E_4$   & $\hspace*{.6mm}E_5$  & $E_6$ & $\hspace*{.6mm}E_7$  & $E_8$  & $\,Z_1$   & $Z_2$  & $Z_3$   & $Z_4$   & $\hspace*{.6mm}Z_5$  
& $Z_6$ & $Z_7$  & $Z_8$  &  \\   \hline \hline
%%\textrm{}& \textit{} & \textit{} & $E_1$   & $E_2$  & $E_3$   & $E_4$   & $E_5$  & $E_6$ & $E_7$  & $E_8$  & $Z_1$   & $Z_2$  & $Z_3$   & $Z_4$   & $Z_5$ 
%% & $Z_6$ & $Z_7$  & $Z_8$  &  \\   \hline \hline
\multicolumn{1}{|l|}{\textit{Total}}
%\textit{Total}  
& \textrm{114}& \textrm{114} & \textrm{9}     & \textrm{110}    & \textrm{59}    & \textrm{99}     & \textrm{59}   & \textrm{109}   & \textrm{87}   & \textrm{114}    & \textrm{0}     & \textrm{114}    & \textrm{109}   & \textrm{114}    & \textrm{83}   & \textrm{114}   & \textrm{104}  & \textrm{114}    & \textrm{\,202.5}    
\\ \hline 
\end{tabular}
\end{center}
\label{tab:exper}
\end{table}

Table~\ref{tab:exper} shows that, for each verification task, the transformation of the CHCs allows 
a very significant
improvement of the performance of the Z3 solver and also an overall
improvement of the Eldarica solver (notably for {\it sat\/} properties).
%%an overall 
%%%a significant 
%%improvement of the performance of the Eldarica solver and a very significant
%%improvement of the Z3 solver.

In particular, before CHC transformation,
Z3 did not prove any of the 114 {\it sat\/} properties of our benchmark.
%\comment{(12 of them were reported `{\it unknown}' within a few seconds). DELETE?}
After CHC transformation, Z3 proved 109 of them to be {\it sat}
(see columns $Z_{1}$ and $Z_{3}$ of Table~\ref{tab:exper}). 
{The time cost of this improvement is very small.
Indeed, most CHC transformations 
take well below~1.5s and only one of them takes a little more than~2s (for details,
see column $T$, where each entry is the sum of the times taken for
the individual transformation tasks of each row).
The times taken by the solvers after transformation (not shown in Table~\ref{tab:exper}) are 
usually quite small. In particular, for the 109 properties
proved {\it sat\/} by Z3, the verification time was almost always below 1s. 
Only for 13 of them, it was between 1s and 4s.}
%Z3 % almost always 
%took less than 1s to prove those 109 {\it sat} properties, 
%and only for 13 of them it required  up to 4s. 
%
% SELECT z3wwo_t FROM results WHERE expanswer='sat' AND  z3wwo='sat'
%
% a few more seconds
% a maximum time of
%
%more than 1s (but 
%none of these six required more than 3.5s). 
For the remaining five {\it sat} properties, % which were not proved {\it sat},
Z3 exceeded the timeout limit. 
%The 12 properties
%that have been proved {\it sat} by Z3 before transformation, were also proved 
%{\it sat} after transformation and always in a shorter time. 
%isThe time cost of this improvement is very small. %limited. 

%%(see 
%%column $T$ where each entry is the sum of the times taken for the verification tasks of each row).

%% SELECT transf_t FROM results

Out of the 
114 {\it sat\/} properties, Eldarica proved 9 {\it sat} properties (all relative to list size) before transformation and
59 {\it sat} properties (relative also to  catamorphisms different from list size) after transformation (see columns $E_{1}$ and $E_{3}$). However, one property that
was proved {\it sat} before transformation, was not proved {\it sat}
%reported \comment{ {\it unknown} } 
after transformation. 
% -- SELECT * FROM results WHERE expanswer='sat' AND eldsrc=expanswer AND eldwwo!=expanswer
This is the only example where the built-in \textit{size} 
function of Eldarica has been more effective than our transformation-based approach.
%\comment{(maybe this is due to the fact that
%Eldarica uses for data structures the default catamorphism {\it size}). 
%Any support to this claim?}

Given the~114~{\it unsat\/} properties, Z3 proved  all of them to be {\it unsat\/} before transformation 
and also after transformation (see columns $Z_{2}$ and $Z_{4}$).
%\comment{ All those proofs almost always took well below~1s 
%(for two of them Z3 took, before transformation,
%3s and 1.8s, respectively). Almost always the proofs after 
%transformation took shorter time with respect to the same proofs 
%before transformation.}
The proofs before transformation took well-below~1s in almost all examples, and after 
transformation took an equal or shorter time for more than half of the cases.
% SELECT ex, z3src_t, z3wwo_t,  ROUND(z3wwo_t/z3src_t, 2)  FROM results WHERE z3src=expanswer
%For the 114 {\it unsat} properties of our benchmark,

Given the~114~{\it unsat} properties, Eldarica proved 110 of them to be {\it unsat} 
before transformation, and only 99 of them after transformation 
(see columns $E_{2}$ and $E_{4}$). 
This is the only case where we experienced a degradation of
 performance after transformation. 
% SELECT ex, eldsrc, eldsrc_t, eldwwo, eldwwo_t FROM results WHERE eldsrc=expanswer AND  expanswer='unsat'
%In general, Eldarica takes longer 
%than Z3 to show that {\it unsat} properties are indeed {\it unsat}.
This degradation may be related to the facts that: %for every property, 
(i)~the number of clauses 
in the transformed set $T_{\mathit{w\&wo}}$ is larger than the number 
%Mau +++ about twice as large as the number
of clauses in the source set, and (ii)~the clauses in~$T_{\mathit{w\&wo}}$
have often more atoms in their bodies with respect to the source clauses. 

%%% Fabio:
%%\comment{
%%This degradation may be related to the fact that 
%%the number and non-linearity of clauses (that is, the number of atoms in their body)
%%for $T_{\mathit{w\&wo}}$ is usually much larger than for the source clauses.
%%}

%If , for every property, 
If we consider the set $T_{\mathit{w}}$, instead of $T_{\mathit{w\&wo}}$, % of transformed
we have a significant decrease in the number of clauses % (about one half) 
and the number of atoms in the bodies of clauses. %(away: about one third). 
In this case, Z3 proved 83 properties to be {\it sat}  (less than for $T_{\mathit{w\&wo}}$, see columns~$Z_3$ and~$Z_5$)
and all 114 properties to be {\it unsat} (as for all other input sets of clauses, see columns~$Z_2$, $Z_4$, and~$Z_6$).
Eldarica proved 59 properties to be {\it sat} 
(the same as for~$T_{\mathit{w\&wo}}$, see columns~$E_3$ and $E_5$) 
and 109 properties to be {\it unsat} (almost the same as for the source clauses, see columns~$E_2$ and $E_6$).

%%If we consider the set $T_{\mathit{w}}$, instead of $T_{\mathit{w\&wo}}$, % of transformed
%%we have a significant decrease of the number of clauses (about one half) and the number of atoms in their
%%bodies (away: about one third). In this case, Z3 proved 83 properties to be {\it sat}  (less than for $T_{\mathit{w\&wo}}$) (see columns~$Z_3$ and~$Z_5$)
%%and all 114 properties to be {\it unsat} (as for all other input sets of clauses) (see columns~$Z_2$, $Z_4$, and $Z_6$).
%%Eldarica proved 59 properties to be {\it sat} 
%%(the same as for~$T_{\mathit{w\&wo}}$) (see columns $E_3$ and $E_5$) 
%%and 109 properties to be {\it unsat} (almost the same as for the source clauses) (see columns $E_2$ and $E_6$).

%%Eldarica proved 59 properties to be {\it sat} 
%%(the same as for $T_{\mathit{w\&wo}}$) (see columns $E_3$ and $E_5$) 
%%and 109 properties to be {\it unsat} (almost the same as for the source clauses) (see columns $E_2$ and $E_6$),
%%whereas Z3 proved 83 properties to be {\it sat}  (less than for $T_{\mathit{w\&wo}}$) (see columns $Z_5$ and $Z_2$)
%%and all 114 properties to be {\it unsat} (as for all other sets of clauses) (see columns $Z_6$, $Z_4$, and $Z_2$).

\indent
{Finally, %for every property, 
we have considered the set $T_{\mathit{wo}}$, instead of $T_{\mathit{w\& wo}}$. 
For the 114 {\it sat} properties, Eldarica proved 87 of them
(see column~$E_7$), while Z3 proved 104 of them  (see column~$Z_7$). 
For the {\it unsat} properties both Eldarica and Z3 proved all of them (see columns~$E_8$ and~$Z_8$).
However, since $T_{\mathit{wo}}$ computes an overapproximation with respect to $T_{\mathit{w\& wo}}$ 
(and also 
with respect to $T_{\mathit{w}}$), when the solver returns the answer {\it unsat\/}, 
one cannot conclude that the property at hand is indeed {\it unsat}. 
Both solvers, in fact, wrongly classified 
10 {\it sat} properties as {\it unsat}.}

In summary, our experimental evaluation shows that \vericatabs with Z3 
as back-end solver outperforms the other CHC solving tools %and techniques 
we have considered. Indeed, our tool shows much higher effectiveness than the 
others 
%%\comment{\cite{HermenegildoPBL05,Su&11}}
%%\comment{quali sono questi? O cambiamo la frase?} 
when verifying \textit{sat} properties, while it retains the excellent performance of Z3 for \textit{unsat} properties.

\section{Conclusions and Related Work} 
\label{sec:RelConcl}
It is well known that the proof of many program properties can be 
reduced to a proof of satisfiability of sets of CHCs~\cite{Bj&15,DeAngelisFGHPP21,Gurfinkel22}. In order
to make it easier to automatically prove satisfiability, whenever a 
program is made out of many functions, 
possibly recursively defined and
depending on each other, 
it is commonly suggested to provide properties also
for the auxiliary functions that may occur in the program. 
Those extra properties basically play the role of lemmas, which often make the proof of a property of interest much easier.  

We have focused our study on the automatic proof of properties
of programs that compute over ADTs, %in the case 
when these properties can be defined using cata\-morphisms.
In a previous paper~\cite{DeAngelisFPP23a}, 
we have proposed an algorithm for
dealing with a multiplicity of properties of 
the various program functions to be proved at the same
time. 
In this paper, we have investigated
an approach, whereby the auxiliary properties need not be explicitly 
defined, but it is enough to  
indicate the catamorphisms involved in 
their specifications. This leaves to the CHC solver the burden of 
discovering the suitable auxiliary properties needed for the proof of 
the property of interest. 
Thus, this much simpler requirement we make %impose %very much simplifies 
avoids the task of providing all the properties of %a full specification for 
the auxiliary functions occurring in  the program.
%proposed in the previously cited paper. 
However, in principle, the proofs of the properties may become harder for the CHC solver.
%it may make the proofs harder. 
Our experimental evaluation shows that this is not 
the case if we follow a transformation-based approach. 
Indeed, %when properties are specified using catamorphisms, 
the results presented in this paper support the following two-step approach:
(1)~use algorithm~\abs\ proposed here to derive a new, transformed set of CHCs from the given initial set of CHCs that translate the program 
together with its property of interest, and then, (2)~use the Z3 solver with 
{\it global guidance}~\cite{KrishnanCSG20} on the derived set.

We have shown that our approach is a valid alternative to the development of algorithms for extending CHC solvers with special purpose mechanisms that handle ADTs.
In fact, recently proposed approaches extend CHC solvers to the case of CHCs  over ADTs through the use of various %auxiliary 
mechanisms such as:
(i)~the combination with inductive theorem proving~\cite{Un&17},
(ii)~the lemma generation based on syntax-guided synthesis from user-specified templates~\cite{Ya&19}, 
(iii)~the invariant discovery based on finite tree automata~\cite{KostyukovMF21},
and (iv)~the use of suitable abstractions on CHCs with recursively defined
function symbols~\cite{GovindSG22}.
%In particular, Govind \textit{et al.}~\cite{GovindSG22} propose an algorithm 
%based on IC3~\cite{Bra11} for solving CHCs modulo catamorphisms.

One key feature of our 
algorithm \abs~is that it is sound and complete with respect to satisfiability, that is, the transformed set of CHCs is satisfiable if and only if so is the initial one. 
In this respect, our results here improve over previous work~\cite{DeAngelisFPP22}, where 
algorithm \Cata~only preserves soundness, that is, if the transformed set of CHCs is satisfiable, then so is the initial one, while if the transformed set is unsatisfiable, nothing can be inferred for the given set.

%%perform the sound and complete CHC transformation using algorithm \abs~proposed in this paper, and then
%%(ii)~use the CHC Z3 solver with 
%%{\it global guidance}~\cite{KrishnanCSG20}.
%%
%%In conclusion, our recommendation for showing {\it sat}/{\it unsat} properties
%%expressed by catamorphisms
%%is: (i)~to perform the sound and complete CHC transformation using algorithm \abs~proposed in this paper, and then
%%(ii)~use the CHC Z3 solver with 
%%{\it global guidance}~\cite{KrishnanCSG20}.

In our experiments, we have also realized the usefulness of having more 
catamorphisms acting together when verifying a specific property.
For instance, in the case of the quicksort program,
when using the catamorphism \isasorted\ alone, Z3 is unable to show 
(within the timeout of 600s) sortedness 
of the output list, while when using also the catamorphisms {\it listmin} and {\it listmax}, after transformation
Z3  proved sortedness in less than 2s.
We leave it for future work to automatically derive the catamorphisms that are
useful for showing the property of interest, even if they are not strictly necessary for
specifying that property.

%\comment{Our experiments also show the usefulness of the catamorphism {\it size} used
%by default by Eldarica when acting on ADTs. %Algebraic Data Types.
%For instance, before transformation, Eldarica was able to prove 
%that bubblesort and insertionsort preserve the size of the output list. However, it was 
%unable to prove (within the timeout) 
%that the same property holds for REVA (i.e., a program for reversing a list 
%by using an accumulator). Nonetheless, after CHC transformation, 
%Eldarica was, indeed, successful in proving that REVA preserves the size of the output list. 
%This fact is an indication
%that our algorithm~\abs\ we presented in this paper may have some advantages (?) over the  
%Eldarica use of the catamorphism {\it size}.}

Our approach is very much related to 
\emph{abstract interpretation}~\cite{CoC77}, which is a methodology for checking properties by interpreting the program as computing over a given abstract domain.
% that is the range of an abstraction function. 
Catamorphisms can be seen as specific abstraction functions. 
Abstract interpretation techniques have been studied also in the field of logic programming. In particular, the CiaoPP preprocessor~\cite{HermenegildoPBL05} 
%for the Ciao logic programming system 
implements abstract interpretation techniques that use {\em type-based norms}, which are a special kind of integer-valued catamorphisms.
These techniques have important applications in \textit{termination analysis}~\cite{BruynoogheCGGV07} and  {\em resource analysis}~\cite{AlbertGGM20}.

Usually, abstract interpretation is the basis for sound analysis 
techniques by computing an (over-)approximation of the concrete 
semantics of a program, and hence these techniques may find 
counterexamples to the properties of interest that hold in the 
abstract semantics, but that are not feasible in the concrete 
semantics. As already mentioned, our transformation guarantees the 
equisatisfiability of the initial and the transformed CHCs, and hence 
all counterexamples found are feasible in the initial CHCs.

Among the various abstract interpretation techniques, the one which is most related to our verification approach, is the 
so-called {\em model-based abstract interpretation} \cite{GallagherBS95}. This
abstract interpretation technique is based 
on the idea of defining a {\em pre-interpretation}, that is, an interpretation of the function symbols of a logic 
program over a specified domain of interest. That pre-interpretation is used for generating, via {\em abstract 
compilation} \cite[Sec. 4.3]{DeAngelisFGHPP21}, a {\em domain program} whose least model is an abstraction 
of the least model of the original program. 
Then, program properties can be inferred from the model of the domain program. One similarity is 
that pre-interpretations of ADT constructors can be seen as catamorphisms. Actually, our definition of a 
catamorphism is more general than the one of a pre-interpretation, in that: (i) we admit non-ADT additional 
parameters as, for instance, in the \textit{listcount} predicate of our introductory example, and (ii) we  
allow mutually dependent predicates in the definitions of catamorphisms.
Another similarity is that the abstract compilation used by model-based abstract interpretation can be seen as a 
program transformation and, indeed, it can be implemented by partial evaluation.
%%, as both the source program 
%%and the target program are logic programs that use possibly different domains. 
%(possibly  +++, even if, possibly, computing on different domains.
However, as already mentioned for other abstract interpretation techniques, that transformation does not guarantee 
equisatisfiability and  by using it, one can prove the satisfiability of the original set of clauses,
but not its unsatisfiability.
%in fact, only the satisfiability of the original set of clauses can be proved using it.
%It should also be noted that the transformation used by model-based abstract interpretation is simpler than \abs, and thus 
%it would make sense to investigate, in future work, whether it can be enhanced so as to achieve the same effect as \abs, 
%possibly in an \textit{ad-hoc} more efficient way.

Our transformation-based approach is, to a large extent, parametric %agnostic 
with respect to the theory of constraints used in the CHCs. 
Thus, it can easily be extended to theories different from \textit{LIA} and \textit{Bool} used in this paper,
and in particular, to other theories such as linear real/rational arithmetic or bit-vectors, as far as they are supported by 
the CHC solver. This is a potential advantage with respect to those abstract interpretation techniques that require the 
design of an ad-hoc abstract domain for each specific program analysis.

\subsubsection*{Acknowledgements} 
We thank Arie Gurfinkel for helpful suggestions on the use of the Z3 (SPACER) solver. We also thank John Gallagher and the anonymous referees of LOPSTR~2023 for helpful comments on previous versions of the paper.
Finally, we express our gratitude to Robert Gl\"uck and Bishoksan Kafle for inviting us to write this improved,
extended  
version of our LOPSTR~2023 paper. The authors are members of the INdAM Research Group GNCS.

\vspace*{2mm}
\noindent
\textit{Competing interests: The authors declare none.}

%% Bibliography
%\bibliographystyle{acmtrans} % ok Manuel Carro: acmtrans instead of tlplike
%\bibliographystyle{tlplike}
%\bibliographystyle{abbrv}
%\bibliographystyle{splncs04} %commented Alberto 2023-06-08
%\bibliography{Smc,Transformation} 

\end{document}